%% file: main.tex
\documentclass[sigconf,nonacm]{acmart}
\usepackage{booktabs}
\usepackage{tabularx}
\usepackage{colortbl}
\usepackage{multirow}
\usepackage{amsmath}
\usepackage{array}
\usepackage{enumitem}
\usepackage{graphicx}
\usepackage{tikz}
\usetikzlibrary{arrows.meta,positioning,fit,shapes.geometric,calc}
\usepackage{pgfplots}
\usepgfplotslibrary{groupplots}
\pgfplotsset{compat=1.18}
\usepackage{xcolor}
\usepackage{pifont}
\usepackage{url}
\usepackage{float}
\usepackage{placeins}

\setcopyright{none}
\renewcommand\footnotetextcopyrightpermission[1]{}
\acmConference[Preprint]{Preprint}{August 2026}{}
\acmYear{2026}
\copyrightyear{2026}
\acmDOI{}
\acmISBN{}

\definecolor{prosnavy}{RGB}{45,62,82}
\definecolor{prosblue}{RGB}{62,111,164}
\definecolor{prosgreen}{RGB}{76,128,111}
\definecolor{prosorange}{RGB}{181,124,74}
\definecolor{prospurple}{RGB}{116,96,141}
\definecolor{prosred}{RGB}{163,79,84}
\definecolor{prosgray}{RGB}{244,245,246}
\definecolor{prosgroupgray}{RGB}{235,238,238}
\definecolor{prosheaderblue}{RGB}{224,234,242}
\definecolor{proslightblue}{RGB}{244,247,249}
\definecolor{proslightred}{RGB}{247,238,236}
\definecolor{resultink}{RGB}{34,44,58}
\definecolor{resultblue}{RGB}{15,82,204}
\definecolor{resultbluebar}{RGB}{92,151,214}
\definecolor{resultnavy}{RGB}{0,45,130}
\definecolor{resultteal}{RGB}{0,115,140}
\definecolor{resultgold}{RGB}{214,126,0}
\definecolor{resultcoral}{RGB}{211,22,22}
\definecolor{resultredbar}{RGB}{226,132,132}
\definecolor{resultgrid}{RGB}{224,228,231}

\pgfplotsset{
  prosplot/.style={
    axis background/.style={fill=white},
    axis lines=left,
    axis line style={draw=resultink!65,line width=.55pt},
    tick style={draw=resultink!55,line width=.45pt},
    tick label style={font=\sffamily\scriptsize,text=resultink},
    label style={font=\sffamily\scriptsize,text=resultink},
    title style={font=\sffamily\bfseries\small,text=resultink},
    grid style={draw=resultgrid,line width=.35pt},
    legend style={font=\sffamily\scriptsize,draw=none,fill=none},
    every axis plot/.append style={line cap=round,line join=round}
  }
}

\newcommand{\PROS}{\textsc{PROS}}
\newcommand{\PROSLong}{\textbf{\underline{P}}roactive \textbf{\underline{R}}efinement \textbf{\underline{O}}f \textbf{\underline{S}}cientific Posters}
\newcommand{\PROSBench}{\textsc{PROS-Bench}}
\newcommand{\checkmarkcell}{\ding{51}}
\newcommand{\xmarkcell}{\ding{55}}
\newcommand{\partialcell}{$\triangle$}
\newcommand{\numPapers}{120}
\newcommand{\numPtwoP}{120}
\newcommand{\numPosterGen}{120}
\newcommand{\numHuman}{40}
\newcommand{\numConference}{40}
\newcommand{\numConferenceNative}{4}
\newcommand{\numConferenceConverted}{36}
\newcommand{\numPosters}{320}
\newcommand{\numNativeOrigin}{164}
\newcommand{\numPDFDerived}{156}
\newcommand{\numEvalPapers}{40}
\newcommand{\numEvalPosters}{120}

\newcommand{\numInstructions}{720}
\newcommand{\numCalibrationPapers}{8}
\newcommand{\numCalibrationPosters}{16}
\newcommand{\PrimaryBackbone}{GPT-5.6 Terra (\texttt{gpt-5.6-terra})}
\newcommand{\InstructionGenerator}{Gemini 3.6 Flash (\texttt{gemini\allowbreak-3.6\allowbreak-flash})}
\newcommand{\PrimaryJudge}{Gemini 3.1 Pro Preview (\texttt{gemini-3.1-pro-preview})}
\input{generated/final_n40_metrics}

\input{generated/final_completed_primary_metrics}

\input{generated/cross_model_metrics}
\input{generated/conference_challenge_metrics}

\begin{document}
\raggedbottom

\title[Beyond Instruction-Driven Editing for Scientific Posters]{Beyond Instruction-Driven Editing: Source-Grounded Problem Discovery with User-Governed Repair for Scientific Posters}

\author{Xingda Lyu}
\authornote{Corresponding author: xlyu18@uw.edu.}
\affiliation{%
  \department{Information School}
  \institution{University of Washington}
  \city{Seattle}
  \state{Washington}
  \country{USA}
}

\author{Honglin Lu}
\authornote{Equal contribution; authors are listed alphabetically by family name.}
\affiliation{%
  \department{Paul G. Allen School of Computer Science \& Engineering}
  \institution{University of Washington}
  \city{Seattle}
  \state{Washington}
  \country{USA}
}

\author{Xinye Luo}
\authornotemark[2]
\affiliation{%
  \department{Department of Applied Mathematics}
  \institution{University of Washington}
  \city{Seattle}
  \state{Washington}
  \country{USA}
}

\author{Shiqi Yang}
\authornotemark[2]
\affiliation{%
  \department{Department of Applied Mathematics}
  \institution{University of Washington}
  \city{Seattle}
  \state{Washington}
  \country{USA}
}

\renewcommand{\shortauthors}{Lyu et al.}

\begin{abstract}
Interactive editors usually assume that users already know what to change. Yet an important interaction state comes earlier: a user may recognize that an artifact is not working without knowing what intervention to request. We call this the \emph{articulation gap}. We introduce \PROS{} (\PROSLong{}), which separates \emph{epistemic initiative} from \emph{behavioral authority}: the system can surface source-grounded candidate problems, while users decide which become repair goals and whether resulting changes are committed. Accepted issues hand off to native-object PPTX editing with validation and reversible preview. We also introduce \PROSBench{}, a source-linked collection of \numPapers{} papers and \numPosters{} editable PPTX posters, including a \numEvalPosters{}-poster matched primary core and a separate conference representation challenge. On the primary core, \PROS{} achieves a mean VLM-rated stage-balanced diagnosis quality score of \FinalETwoQuality{} on a 0--100 scale and \FinalEThreeRate\% operator-verified target resolution among accepted diagnoses. Temporally blinded automated scoring yields a +\FinalEFourUplift{}-point paper-macro accepted-target uplift, yet \FinalEFourNegative\% of assessable accepted targets decline. This divergence shows why problem discovery, local resolution, and realized outcome should be evaluated separately. More broadly, intelligent editors can support problem discovery before a concrete edit request exists without taking authority over consequential change.
\end{abstract}



\maketitle

\input{figures/fig1_system_overview}

\section{Introduction}

Scientific posters are unusually dense communication artifacts. A single page must compress a research paper while coordinating text, figures, captions, reading order, hierarchy, whitespace, and visual emphasis. Recent systems have made substantial progress in generating posters from papers \cite{pang2025paper2poster,sun2026p2p,zhang2026postergen}, and interactive agents increasingly support natural-language editing of posters, slides, and layered visual documents \cite{shi2026apex,jung2026talkslides,ofengenden2025pptarena,lin2026mildedit}. Artifact construction and request execution are becoming steadily more capable, yet many editing interfaces still begin from one implicit assumption: the user already knows what should change.

The hard part can arise before execution. An author may know that a poster is difficult to follow, visually crowded, or scientifically unconvincing without knowing whether the underlying problem is structural, compositional, or communicative. This is not instruction disambiguation: an adequate edit instruction may not yet exist. We call the distance between sensing that an artifact is not working and formulating a localized, actionable intervention the \emph{articulation gap}. This gap reflects epistemic incompleteness: progress depends on recognizing what is missing from consideration \cite{kaur2026knowing}. Computational critiquing research similarly argues that design tools can reveal implicit artifact features, not only correct defects that users already know how to name \cite{fischer1993critics,nakakoji1998talkback}.

An intelligent interface need not choose between purely reactive assistance and autonomous intervention. We separate \emph{problem discovery} from \emph{consequential change}. Users retain a direct path for intentional edits; when the right intervention is still unclear, an explicitly invoked diagnosis path can surface inspectable, evidence-linked issues. Direct editing executes articulated intent, whereas diagnosis-guided refinement helps form it. This framing builds on mixed-initiative work that combines user steering with automated services \cite{horvitz1999mixed} and on grounded accounts of when intervention is warranted \cite{kaur2026knowing}.

We call the resulting system \PROS{}: \emph{Proactive Refinement of Scientific Posters}. It operationalizes \emph{bounded epistemic initiative}: proactive does not mean unsolicited interruption or autonomous modification. Users invoke diagnosis, decide which candidate issue becomes a repair goal, inspect the preview, and choose whether to commit it. Direct requests and diagnosis-guided refinement share the same source-grounded artifact state and repair substrate (Figure~\ref{fig:overview}).

The system couples a replaceable reasoning model with deterministic detector and artifact services that construct the editable scene, enforce typed operations, execute PPTX changes, and validate candidates. A representation policy distinguishes structured, freeform, and uncertain layouts: reliable structure permits coordinated reflow, whereas uncertain geometry receives conservative local intervention. We characterize the assembled workflow across replaceable backbones; a foundation model alone is not treated as an equivalent editing interface.

Our central question is whether an intelligent editor can help before a concrete edit instruction exists while keeping consequential change under user control. We test the diagnosis-guided path at three distinct points: whether \PROS{} surfaces grounded candidate problems, turns accepted issues into executable repairs, and improves the accepted target condition. E1 separately verifies that the shared executor can carry out articulated requests. Questions of effort, experienced control, and user preference require a controlled user study and remain outside this artifact-centered evaluation.

We evaluate the workflow with \PROSBench{}, which preserves the relationship among source paper, editable PPTX, and rendered poster. The same 40 papers supply Paper2Poster, PosterGen, controlled expert-authored, and public conference-poster variants. The first three form the primary repair core; conference artifacts remain a separate representation challenge. This pairing exposes unequal repair opportunity instead of assuming that every starting artifact needs intervention. The evaluation follows the interaction path: explicit-request execution (E1), problem discovery (E2 diagnosis), accepted-issue repair (E3), and automated accepted-target outcome judged blind to temporal order (E4).

We investigate three research questions:
\begin{description}[leftmargin=0.58in,style=nextline,topsep=-2pt]
\item[RQ1: End-to-end capability.] How effectively does \PROS{} execute articulated requests, surface useful candidate issues before a concrete edit instruction exists, repair protocol-accepted diagnoses, and improve accepted target conditions?
\item[RQ2: Discovery-to-outcome propagation.] Where do losses arise along the discovery--governance--repair--outcome trajectory, from E2 problem discovery through governance and E3 repair to E4 realized outcome?
\item[RQ3: Intervention calibration.] How do diagnosis yield, non-intervention patterns, and repair success vary across task stage, poster provenance, and representation conditions that expose different repair opportunities?
\end{description}
We additionally probe backbone sensitivity on a fixed eight-paper subset as an exploratory portability analysis, not as a model ranking.

This work contributes:
\begin{enumerate}[leftmargin=*]
    \item \textbf{A governed model of proactive editing} that separates epistemic initiative from behavioral authority, enabling inspectable problem discovery before a concrete edit instruction exists while reserving goal adoption and commitment for the user.
    \item \textbf{\PROS{}, a working operationalization} that connects user-invoked, source-grounded diagnosis to semantic click-to-reference interaction, explicit issue acceptance, native PPTX grounding, temporary execution, validation, and reversible preview; its repair authority contracts when representation reliability is lower.
    \item \textbf{\PROSBench{} and a capability decomposition} spanning \numPapers{} source papers and \numPosters{} editable posters. The benchmark contains 40 source-matched groups, a separate conference representation challenge, and 80 additional generated pairs. Its evaluation separately measures explicit-request execution, problem discovery, accepted-issue repair, and realized accepted-target outcome over all \numEvalPosters{} primary-core posters and \numInstructions{} frozen instructions.
\end{enumerate}

\section{Related Work}

\subsection{Scientific Poster Generation and Editable Scientific Artifacts}

Scientific-poster research has historically emphasized generation, layout understanding, and scarce paired data. The NJU--Fudan Paper--Poster dataset paired papers with panel-annotated posters for generation research \cite{qiang2017posterpaper}; SciPostLayout provides 7,855 posters with manual layout annotations and a smaller set of paper--poster pairs \cite{tanaka2024scipostlayout}; SciPostLayoutTree adds reading-order and parent--child relations \cite{tanaka2025scipostlayouttree}; SciPostGen studies paper-conditioned layout generation \cite{inadumi2025scipostgen}; and PosterIQ broadens evaluation to composition, typographic hierarchy, semantic intent, and design quality \cite{feng2026posteriq}. These resources establish that poster quality depends on both structure and communication, but raster, PDF, or layout-only representations cannot by themselves evaluate native-object targeting, preservation, and executable repair.

Recent systems generate structured posters directly from papers. Paper2Poster produces editable PPTX output through parsing, layout planning, and iterative visual feedback \cite{pang2025paper2poster}; P2P uses specialized generation roles and source-specific checks \cite{sun2026p2p}; PosterGen coordinates content, layout, style, and rendering roles \cite{zhang2026postergen}; and PosterForest introduces a hierarchical Poster Tree with coordinated content/layout refinement \cite{choi2026posterforest}. Any2Poster broadens source modalities and uses iterative visual feedback \cite{vinaykumar2026any2poster}. PosterHarness makes generation constraints auditable but studies a 12-paper placeholder-first generation regime rather than post-hoc editable-artifact refinement \cite{yang2026posterharness}. Concurrent PosterMELD emphasizes controllable design diversity, editable print-ready PPTX output, and deterministic/VLM gates during generation \cite{hu2026postermeld}. Adjacent systems such as AutoFigure, AutoFigure-Edit, and LiveFigure likewise demonstrate growing interest in structured and editable scientific illustrations \cite{zhu2026autofigure,lin2026autofigureedit,shao2026livefigure}. These works show that editable scientific outputs are emerging, not absent. Their primary task is construction or generation-time quality control; \PROS{} instead begins from an existing editable poster and asks what should be noticed, what should be changed, and how to preserve the artifact while doing so.

\begin{table*}[t]
\caption{Positioning across poster generation, structured artifact editing, and computational critique. This table positions task and interaction scope; it is not a quality ranking. ``User edit'' means editing the target artifact rather than only critique text; ``Pre-inst. diag.'' means diagnosis before a concrete edit instruction. $\triangle$ denotes partial support or generation-time/internal repair rather than the full capability.}
\label{tab:related-systems}
\centering
\scriptsize
\setlength{\tabcolsep}{2.0pt}
\renewcommand{\arraystretch}{0.98}
\arrayrulecolor{resultink}
\begin{tabular}{p{0.205\linewidth}*{7}{>{\centering\arraybackslash}p{0.084\linewidth}}}
\toprule
\color{resultink}\textbf{System} & \color{resultink}\textbf{Existing} & \color{resultink}\textbf{Source} & \color{resultink}\textbf{User edit} & \color{resultink}\textbf{Pre-inst. diag.} & \color{resultink}\textbf{Native repair} & \color{resultink}\textbf{Verify} & \color{resultink}\textbf{Accept gate} \\
\midrule
\rowcolor{prosgroupgray}\multicolumn{8}{c}{\color{resultink}\textit{Scientific-poster generation}}\\
Paper2Poster \cite{pang2025paper2poster} & \xmarkcell & \checkmarkcell & \xmarkcell & \xmarkcell & \partialcell & \partialcell & \xmarkcell \\
P2P \cite{sun2026p2p} & \xmarkcell & \checkmarkcell & \xmarkcell & \xmarkcell & \partialcell & \partialcell & \xmarkcell \\
PosterGen \cite{zhang2026postergen} & \xmarkcell & \checkmarkcell & \xmarkcell & \xmarkcell & \partialcell & \partialcell & \xmarkcell \\
PosterForest \cite{choi2026posterforest} & \xmarkcell & \checkmarkcell & \xmarkcell & \xmarkcell & \partialcell & \partialcell & \xmarkcell \\
Any2Poster \cite{vinaykumar2026any2poster} & \xmarkcell & \checkmarkcell & \xmarkcell & \xmarkcell & \partialcell & \partialcell & \xmarkcell \\
PosterMELD \cite{hu2026postermeld} & \xmarkcell & \checkmarkcell & \xmarkcell & \xmarkcell & \partialcell & \partialcell & \xmarkcell \\
\rowcolor{prosgroupgray}\multicolumn{8}{c}{\color{resultink}\textit{Instruction-driven editing of structured artifacts}}\\
APEX \cite{shi2026apex} & \checkmarkcell & \checkmarkcell & \checkmarkcell & \xmarkcell & \checkmarkcell & \checkmarkcell & \partialcell \\
Talk-to-Your-Slides \cite{jung2026talkslides} & \checkmarkcell & \xmarkcell & \checkmarkcell & \xmarkcell & \checkmarkcell & \xmarkcell & \xmarkcell \\
PPTPilot / PPTArena \cite{ofengenden2025pptarena} & \checkmarkcell & \xmarkcell & \checkmarkcell & \xmarkcell & \checkmarkcell & \checkmarkcell & \xmarkcell \\
MiLDEdit \cite{lin2026mildedit} & \checkmarkcell & \xmarkcell & \checkmarkcell & \xmarkcell & \checkmarkcell & \partialcell & \xmarkcell \\
SMART-Editor \cite{mondal2026smarteditor} & \checkmarkcell & \xmarkcell & \checkmarkcell & \xmarkcell & \partialcell & \checkmarkcell & \xmarkcell \\
\rowcolor{prosgroupgray}\multicolumn{8}{c}{\color{resultink}\textit{Computational critique and user-governed repair}}\\
DesignScape \cite{odonovan2015designscape} & \checkmarkcell & \xmarkcell & \partialcell & \checkmarkcell & \partialcell & \xmarkcell & \checkmarkcell \\
Scout \cite{swearngin2020scout} & \checkmarkcell & \xmarkcell & \partialcell & \checkmarkcell & \partialcell & \partialcell & \checkmarkcell \\
VizLinter \cite{chen2022vizlinter} & \checkmarkcell & \xmarkcell & \xmarkcell & \checkmarkcell & \checkmarkcell & \partialcell & \partialcell \\
VizCrit \cite{li2026vizcrit} & \checkmarkcell & \xmarkcell & \partialcell & \checkmarkcell & \xmarkcell & \xmarkcell & \checkmarkcell \\
ProactiveVA \cite{zhao2025proactiveva} & \checkmarkcell & \xmarkcell & \partialcell & \checkmarkcell & \partialcell & \partialcell & \checkmarkcell \\
PosterMate \cite{shin2025postermate} & \checkmarkcell & \checkmarkcell & \checkmarkcell & \checkmarkcell & \partialcell & \xmarkcell & \checkmarkcell \\
Agentic-DRS \cite{nag2026agenticdrs} & \checkmarkcell & \xmarkcell & \xmarkcell & \checkmarkcell & \xmarkcell & \xmarkcell & \xmarkcell \\
VLM chart repair \cite{bonas2026chartrepair} & \checkmarkcell & \xmarkcell & \partialcell & \checkmarkcell & \checkmarkcell & \checkmarkcell & \checkmarkcell \\
SlideAudit \cite{zhang2025slideaudit} & \checkmarkcell & \xmarkcell & \xmarkcell & \checkmarkcell & \partialcell & \partialcell & \xmarkcell \\
CritiqueCrew \cite{chen2026critiquecrew} & \checkmarkcell & \xmarkcell & \partialcell & \checkmarkcell & \checkmarkcell & \xmarkcell & \checkmarkcell \\
Criticmate \cite{ko2026criticmate} & \checkmarkcell & \xmarkcell & \xmarkcell & \checkmarkcell & \xmarkcell & \xmarkcell & \checkmarkcell \\
\midrule
\rowcolor{proslightblue}\textbf{\PROS{} (ours)} & \checkmarkcell & \checkmarkcell & \checkmarkcell & \checkmarkcell & \checkmarkcell & \checkmarkcell & \checkmarkcell \\
\bottomrule
\end{tabular}
\arrayrulecolor{black}
\end{table*}

\subsection{Instruction-Driven Editing and Structured Design Evaluation}

Editing differs from generation because a system must preserve object identities, relations, and unrelated content. PPTAgent, PPTC, PPTArena/PPTPilot, Talk-to-Your-Slides, and MiLDEdit establish increasingly capable instruction-driven editing over presentations or layered design documents \cite{zheng2025pptagent,guo2024pptc,ofengenden2025pptarena,jung2026talkslides,lin2026mildedit}. SMART-Editor moves closer to underarticulated editing by translating implicit, cascading instructions into coordinated poster or webpage changes and checking global coherence; it nevertheless begins from a requested outcome rather than deciding what, if anything, warrants intervention \cite{mondal2026smarteditor}. PPT-Eval formalizes rubric-based evaluation for PowerPoint creation/editing tasks and explicitly rewards partial progress while penalizing unnecessary changes \cite{gandhi2026ppteval}. SlideAudit develops an expert-informed taxonomy over 2,400 presentation slides and directly evaluates LLM-based design-flaw detection and remediation, making it the closest diagnosis benchmark outside scientific posters \cite{zhang2025slideaudit}. Broader design benchmarks similarly show that current multimodal models still struggle with precise spatial reasoning, layered editing, typography, and vector manipulation \cite{deganutti2026graphicdesignbench,lin2026graphicdesign}.

APEX is the closest poster-specific editor. It takes an editable poster, associated paper, and user instruction, maps requests to multi-level APIs, and uses a review-and-adjustment loop; APEX-Bench contains 514 human-refined instructions over 59 paper--poster pairs \cite{shi2026apex}. It is an essential precursor, but its scored cases begin from an edit instruction. \PROS{} shares the goal of reliable executable editing while addressing an additional interaction state: the user may know that the poster is not working without yet having a concrete instruction. At the benchmark level, existing resources separately cover poster generation, instruction-driven PPTX editing, or slide-level design critique. Our claim is consequently narrower than ``the first PPTX'' or ``the first poster editor'': in our review, no prior resource combines four source-matched poster provenances with explicit execution, problem discovery before an edit instruction, user-governed source-grounded repair, and automated accepted-target outcome measurement over persistent editable artifacts. Direct editing and diagnosis-guided refinement converge on the same repair executor instead of becoming separate systems.

\subsection{Computational Critique and User-Governed Design Support}

Computational critique has a long history in design tools. Embedded critics use contextual knowledge to detect problematic situations \cite{fischer1993critics}, while representational talkback emphasizes revealing features that may remain implicit to the designer \cite{nakakoji1998talkback}. DesignScape and Scout provide suggestive or mixed-initiative support for layout exploration \cite{odonovan2015designscape,swearngin2020scout}; VizLinter detects and repairs visualization-rule violations \cite{chen2022vizlinter}; and VizCrit studies how awareness- and solution-centered annotations make computational design feedback actionable \cite{li2026vizcrit}. ProactiveVA provides context-aware assistance in visual analytics \cite{zhao2025proactiveva}, while VLM chart-repair systems diagnose design flaws and support selective re-rendering \cite{bonas2026chartrepair}. PosterMate is an especially important poster-design neighbor: audience-persona agents provide component- and theme-level feedback that users may discuss, apply, or further edit, although its target is advertisement posters rather than source-grounded scientific communication \cite{shin2025postermate}. Agentic-DRS uses specialized evaluators to produce multi-facet graphic-design feedback \cite{nag2026agenticdrs}. Most directly, CritiqueCrew turns multi-perspective Figma critique into a user-selected agenda and in-context remediation workflow, and evaluates the interaction in two controlled studies \cite{chen2026critiquecrew}. Criticmate similarly exposes perception, comprehension, and projection as editable intermediate artifacts for stagewise UI co-critique, showing how critique structure can create meaningful points for human correction and contribution \cite{ko2026criticmate}.

Mixed-initiative research emphasizes that user control and automated assistance can coexist rather than replace one another \cite{horvitz1999mixed,deterding2017mixed}. Proactive text-to-image agents address underspecified intent during generation, while PROPER models task-relevant knowledge gaps as selectively activated dimensions in conversational assistance \cite{hahn2025proactivet2i,kaur2026proper}; \PROS{} instead diagnoses deficiencies in an already shared editable artifact. Recent IUI studies further show that assistance delivery mode and intervention timing can affect user perception and receptivity even when task outcomes are similar \cite{luo2026mixedinitiative,kuo2026proactive}. These works motivate a bounded account of proactivity in \PROS{}: the user chooses when to invoke diagnosis, the system takes epistemic initiative only in expanding what may warrant attention, and explicit user acceptance separates a candidate issue from behavioral commitment.

Together, these neighbors sharpen the novelty boundary. Our claim is not that diagnosis, repair, source grounding, structured editing, or user approval is individually unprecedented. It is that \PROS{} connects them across the pre-instruction articulation gap, explicit authority boundaries, and persistent editable artifacts, while \PROSBench{} evaluates the discovery--governance--repair--outcome trajectory.

\subsection{Grounding, Verification, and Scalable Judging}

Source-grounded generation can reduce unsupported content by conditioning outputs on retrieved evidence \cite{lewis2020rag}. Iterative model feedback can improve outputs, but self-critique alone does not guarantee artifact preservation \cite{madaan2023selfrefine}. \PROS{} therefore combines model reasoning with deterministic artifact services and treats internal verification as part of the workflow rather than as the final evaluator.

Open-ended visual quality is also difficult to score automatically. LLM-as-a-judge and multimodal-judge studies report useful agreement with humans but also position, self-preference, hallucination, and consistency biases \cite{zheng2023judge,chen2024mllmjudge}; design-specific evaluations likewise find that multimodal design judgment remains imperfect and multidimensional \cite{lin2026graphicdesign}. We use a frozen high-capacity multimodal judge for scalable full-set E2/E4 scoring and randomize E4 A/B labels before judgment.

\section{Design Principles and Authority Boundaries}

Five principles connect the interaction claim to implementation. \textbf{P1: inspectable grounding.} Diagnoses expose an issue class, target, rationale, and---for scientific content---source evidence. \textbf{P2: assistance without authority.} Users may bypass diagnosis, dismiss or select a proposal, revise the queued request, and accept or reject the candidate artifact; a proposal never becomes a committed change without explicit action \cite{horvitz1999mixed,amershi2019guidelines,luo2026mixedinitiative,kuo2026proactive}. \textbf{P3: editable structure.} Rendered appearance remains linked to addressable native objects, section membership, and evidence rather than treating pixels as the document. \textbf{P4: representation-aware strength.} Structured layouts permit coordinated reflow, while freeform or uncertain layouts receive conservative local edits. \textbf{P5: reversible, failure-contained commitment.} Typed operations, temporary copies, validation, and candidate preview block detectable invalid targets, unsafe geometry, ownership drift, and failed execution before commitment.

\paragraph{Scope.}
\PROS{} targets one-slide editable PPTX scientific posters containing text, figures, captions, tables, shapes, and grouped regions. It does not claim to infer latent author intent, create novel scientific figures, or perfectly recover edit semantics from a flattened PDF. The primary matched evaluation includes native-origin PosterGen and expert-authored artifacts and PDF-derived Paper2Poster artifacts converted with Adobe Acrobat or WPS. Public conference posters---\numConferenceNative{} distributed as native PPTX and \numConferenceConverted{} converted from PDF with Adobe Acrobat---form a separate representation challenge. Provenance, representation origin, converter, diagnosis opportunity, and unsupported-representation outcomes are retained rather than hidden as preprocessing. The diagnosis taxonomy is an operational core, not a complete ontology of poster quality.

\section{The \PROS{} Workflow}

The workflow separates model reasoning from deterministic artifact services and separates diagnosis from repair. The reasoning backbone can be replaced through a canonical adapter; the artifact parser, executor, policy guards, and external evaluator remain fixed across backbone configurations.

\subsection{Shared Artifact State and Layout Profile}

At each refinement step, \PROS{} maintains the source-paper text, current working PPTX, rendered preview, native object inventory, inferred section/scene structure, layout profile, and any temporary candidate artifact. The inventory exposes both user-facing object labels and native shape identifiers. For freeform posters, lightweight markers preserve section ownership across accepted edits; the system does not assume that arbitrary PPTX reconstruction preserves every inferred relation.

The inventory is not a raw list of shape IDs. \PROS{} infers semantic roles from native object type, text, style, geometry, neighborhood, and section ownership; optional model arbitration is used only for ambiguous assignments. It then derives the relations required by the current detectors and executor, including section ownership, same-column and above/below relations, overlap, and title/content/figure companionship. User-facing labels (e.g., ``Figure 2'' or ``Section Content 5'') remain linked to native identifiers. Before a transaction, target mappings are snapshotted so that an insertion or deletion cannot silently retarget a later action. Open-ended repair prompts contain the relevant addressable inventory and extracted source-paper text, not only a rendered image; previews remain available for user inspection and external visual evaluation.

Before repair, the system also assigns a layout profile using section ownership, visual columns, independent panels, title/content grouping, and cross-region geometry. \emph{Structured} posters can use coordinated section reflow. \emph{Freeform} posters receive a stable section registry and local safeguards; automatic global reflow is disabled, and a candidate can be rejected if local edits break the authored geometry. \emph{Uncertain} posters likewise disable aggressive global reflow. Representation reliability bounds the executor's geometric authority.

\subsection{Semantic Reference and Request Composition}

The rendered poster remains an interaction surface, not a passive screenshot. Hovering over a semantically addressable region reveals its label and bounds; clicking appends that label to the shared repair-request box. This click-to-reference technique lets users localize a direct request without memorizing a shape ID, while the executor resolves the label against native PPTX state. The same box receives the repair text of an accepted diagnosis, which users may revise before planning. Users may also upload raster assets and request supported add, replace, or delete operations. Semantic pointing, free-form language, and diagnosis-generated requests therefore converge on one typed action interface instead of becoming separate editing subsystems.

\subsection{Two Entry Paths: Intent Execution and Intent Formation}

\paragraph{Direct reactive editing.} The user supplies an explicit request, optionally composed with a clicked semantic label. Deterministic request patterns first take a rule-based fast path; other requests are passed to the model planner with relevant addressable artifact state, extracted source context, uploaded-asset names when applicable, and the typed action schema. This hybrid path preserves exact mappings for common local operations while retaining open-ended planning. Full diagnosis is unnecessary unless the request itself is ambiguous; the render remains visible to the user and is re-generated for preview and evaluation.

\paragraph{Diagnosis-guided refinement.} The user invokes diagnosis without specifying an edit operation. The system proposes candidate issues for inspection. Each issue carries a class, target, rationale, severity information, and a source heading or source-derived repair payload when scientific content is implicated. The user may dismiss or accept it. Acceptance only queues a repair request, which remains editable before planning. If the user leaves that request unchanged, the detector's typed operation plan is used; if the user rewrites it, the request is replanned through the direct path. Both routes use the same executor, temporary preview, and explicit commit boundary.

The asymmetry between these paths is deliberate. Direct editing begins from articulated intent; diagnosis-guided refinement supports intent formation by proposing what may deserve attention. Both remain user initiated, and both require user approval before a candidate artifact replaces the current one.

\subsection{Three Diagnostic Layers in Runtime Order}
\label{sec:three-stage-diagnosis}

\PROS{} visits three user-visible stages in runtime order: \emph{Stage 1 structural validity $\rightarrow$ Stage 2 scientific communication $\rightarrow$ Stage 3 spatial composition}. Exactly one stage is active at a time; users may accept several issues within that stage, which are deduplicated and composed in the shared request box. The structural stage is a standing safety gate: it is rechecked after accepted edits and regains focus if a candidate introduces a collision, overflow, or related mechanical failure. A cleared focus stage is locked before the workflow advances, preventing stale diagnoses from reopening it. The order is operational: scientific-content changes may alter text quantity or section demand, so spatial balancing is most useful after communication-level changes have stabilized.

\begin{enumerate}[leftmargin=*]
    \item \textbf{Structural validity.} Detects implemented mechanical failures that make editing unsafe or readability invalid: collisions, occlusion, overflow, and out-of-bounds placement. Evidence is primarily native geometry, text bounds, addressable objects, and render checks, consistent with linter-style and structured-editing work \cite{chen2022vizlinter,jung2026talkslides}.
    \item \textbf{Scientific communication.} Checks whether the poster communicates the source paper with sufficient coverage and focus. It surfaces omitted results or methods, redundancy/overload, metadata noise, obvious section-role mismatch, and source-coverage inconsistency. This layer is conditioned on extracted source-paper content but does not claim exhaustive scientific fact-checking; poster-generation work likewise shows that visual plausibility alone does not guarantee scientific coverage \cite{pang2025paper2poster,sun2026p2p,vinaykumar2026any2poster}.
    \item \textbf{Spatial composition.} Assesses the final arrangement of valid, communication-ready content: whitespace allocation, alignment, section balance, grouping, hierarchy, and reading order. Prior visual-authoring and poster studies motivate treating balance, proximity, saliency, and reading order as distinct poster-level concerns \cite{odonovan2015designscape,swearngin2020scout,tanaka2025scipostlayouttree,feng2026posteriq,faulkes2023wall,galibourg2026visual}.
\end{enumerate}

The stages form an operational scaffold rather than a claim that poster quality decomposes into independent dimensions.

\begin{table*}[t]
\caption{Operational diagnosis taxonomy. The six classes define the current system and benchmark scope, not an exhaustive theory of scientific-poster quality.}
\label{tab:taxonomy}
\centering
\small
\arrayrulecolor{resultink}
\begin{tabularx}{\linewidth}{p{0.25\linewidth}X p{0.28\linewidth}}
\toprule
\color{resultink}\textbf{Layer / class} & \color{resultink}\textbf{Representative deficiencies} & \color{resultink}\textbf{Primary evidence} \\
\midrule
\rowcolor{prosgray}\multicolumn{3}{c}{\color{resultink}\textit{Structural validity}}\\
Collision & figure--text or text--text overlap; occlusion; unsafe z-order interaction & native geometry, overlap masks, IDs \\
Readability & overflow; clipped or out-of-bounds text; invalid container extent & text bounds, layout checks, render \\
\rowcolor{prosgray}\multicolumn{3}{c}{\color{resultink}\textit{Scientific communication}}\\
Content sufficiency & omitted method, result, limitation, or necessary evidence & poster content, source-paper evidence \\
Content alignment / noise & redundancy, overload, metadata leakage, obvious role mismatch, source-coverage inconsistency & section roles, source alignment \\
\rowcolor{prosgray}\multicolumn{3}{c}{\color{resultink}\textit{Spatial composition}}\\
Whitespace & excessive, fragmented, or poorly allocated blank regions & region geometry, relative blank area \\
Layout balance / hierarchy & weak alignment/grouping, section imbalance, unclear emphasis or reading order & visible layout relations, grouping \\
\bottomrule
\end{tabularx}
\arrayrulecolor{black}
\end{table*}

\subsection{Accepted-Issue Handoff and Repair}

Diagnosis and repair remain separate. A user instruction or accepted diagnosis is converted into a natural-language repair request, after which the planner chooses from the shared typed operation interface. Multiple action sequences may satisfy the same request; evaluation therefore focuses on outcome constraints and collateral damage rather than exact imitation of one human action trace.

Scientific-content planning is conditioned on extracted source evidence, and content diagnoses expose a source heading or source-derived repair payload when available. This grounding constrains generation but is not treated as a proof of factual correctness: users inspect the proposed text before commitment, and E3 externally verifies whether the accepted target is resolved. The runtime therefore claims evidence-conditioned planning and governed review, not automatic scientific verification.

The executor applies candidate actions to a temporary PPTX, resolves targets against a transaction snapshot of the current object inventory, reopens the deck, and re-renders when needed. Structured layouts may invoke coordinated reflow. Freeform execution is fail-closed: it verifies section identity and ownership, rejects unexpected non-target mutation or new collisions, and never enables acceptance for an invalid candidate. Failed multi-operation batches retry actions individually so one failure does not discard safe batch-mates; rejected text additions may retry shorter variants before abstaining. The UI exposes the candidate render, operations, issue/evidence context, and warnings while retaining the current artifact unchanged. Acceptance commits the candidate; rejection discards it. A central controller owns context, stage transitions, artifact state, and commitment permissions, keeping the workflow auditable across replaceable reasoning backbones.

\section{\PROSBench{}}

\subsection{Benchmark Composition}

A refinement benchmark must link the source paper, editable poster, and render. Image-only corpora cannot test native-object targeting or preservation, while PDF-to-PPTX conversion introduces representation errors. Existing resources cover poster generation, layout analysis, or instruction-first editing, but none in our review couples these three artifacts, four source-matched provenances, and the discovery--governance--repair--outcome trajectory. \PROSBench{} accordingly treats provenance, representation origin, and converter as benchmark variables.

\input{figures/fig2_benchmark}

\PROSBench{} contains \numPapers{} unique open-access papers from CVPR (35), ICML (31), NeurIPS (28), and ICLR (26), spanning 2023--2026 and six broad topic families. Paper2Poster and PosterGen process each paper independently. We convert all \numPtwoP{} Paper2Poster PDFs to editable PPTX using Adobe Acrobat or WPS and retain the PDF, converter identity, and derivative; PosterGen contributes \numPosterGen{} native-origin PPTX outputs. From the same pool, \numHuman{} papers are assigned to expert creators---graduate students or researchers experienced in scientific papers, presentations, or posters---who construct one source-matched native PPTX without inspecting either generated variant. These posters use one shared 48~$\times$~36-inch public PosterNerd template \cite{posternerdtemplates}, which controls low-level structure while leaving paper-specific decisions about content, emphasis, figures, and organization to the creator.

The benchmark contains \numPosters{} editable starting posters: 40 four-way source-matched groups (160 posters) and 80 generated pairs (160 posters). Across the collection, \numNativeOrigin{} artifacts are native-origin PPTX and \numPDFDerived{} are PDF-derived. The primary E1--E4 study uses the \numEvalPosters{} Paper2Poster, PosterGen, and expert-authored posters from the 40 matched papers. Their conference variants remain source-matched but form a separate representation challenge: mature design and heterogeneous recovered editability create a different repair-opportunity condition. The remaining 80 generated pairs extend breadth but do not enter the headline estimates.

All artifacts receive stable identifiers, hashes, open/render checks, object-inventory extraction, slide-dimension checks, and provenance validation. No component is trained on \PROSBench{}, and a disjoint calibration set is excluded from reported estimates. Appendix~\ref{app:benchmark-details} documents creator QA, acquisition and conversion, calibration, manifests, and the boundary between retained and unavailable metadata.

\begin{table*}[t]
\caption{Full \PROSBench{} composition and empirical role. Forty source papers form a 120-poster three-variant primary repair core plus a 40-poster conference representation challenge; the remaining 80 papers form a 160-poster generated-pair breadth extension.}
\label{tab:dataset}
\centering
\small
\arrayrulecolor{resultink}
\begin{tabular}{lrrl}
\toprule
\color{resultink}\textbf{Poster provenance} & \color{resultink}\textbf{Source papers} & \color{resultink}\textbf{Posters} & \color{resultink}\textbf{Role} \\
\midrule
Paper2Poster & 120 & 120 & 40 primary + 80 breadth; PDF-derived \\
PosterGen & 120 (same) & 120 & 40 primary + 80 breadth; native-origin \\
Expert-authored (shared template) & 40 (subset) & 40 & primary low-headroom condition; native-origin \\
Conference poster & 40 (same subset) & 40 & separate challenge; 4 native + 36 converted \\
\midrule
\textbf{Total} & \textbf{120} & \textbf{320} & \textbf{40 four-way + 80 two-way groups} \\
\bottomrule
\end{tabular}
\arrayrulecolor{black}
\end{table*}

\subsection{E1 Instruction Construction}

For each primary-core poster, \InstructionGenerator{} receives the frozen render and matched paper and drafts six source-linked tasks: two Structural, two Scientific, and two Spatial. An author validates necessity, localization, executability, non-duplication, preservation, and scientific support before any evaluated-backbone execution. Drafts are retained verbatim, boundedly revised, or replaced only within the same stage and slot when already satisfied, duplicate, unsupported, unlocatable, or out of scope. The release preserves both original and frozen instructions, reasons for intervention, expected outcomes, and protected constraints; no task is removed based on execution outcome. Full decision rules appear in Appendix~\ref{app:benchmark-details}.

\subsection{Evaluation Assets for E2--E4}

E2 stores and scores each predicted issue rather than assuming an exhaustive issue list. E3 links every operator-accepted diagnosis, repair request, and verified target-resolution outcome through a stable issue ID. E4 compares original and final posters under randomized opaque A/B labels for the accepted target condition, using source evidence for Scientific items. This design keeps discovery quality, governed repair, and realized target uplift distinct; it does not claim holistic aesthetics or publication readiness.

\subsection{Release Structure and Auditability}

The release separates \texttt{system/}, \texttt{dataset/}, \texttt{evaluation/}, and \texttt{results/}, keeping fixed task definitions apart from outputs. It preserves diagnosis text, stable IDs, governance decisions, repair requests, E1/E3 outcomes, judge responses, evidence manifests, captured hashes, and available before/final artifacts. Missing historical telemetry or fine-grained traces remain blank rather than reconstructed; Appendix~\ref{app:reproducibility} specifies the released fields and evidence boundaries.

\section{Evaluation}

\subsection{Evaluation Perspective: Characterizing a Complete Interactive System}

We evaluate \PROS{} as an interactive workflow rather than a model call. Its replaceable reasoning backbone operates within synchronized native-PPTX state, semantic targets, typed actions, representation-aware execution, governed handoff, validation, and reversible commitment. E1--E4 therefore measure explicit-request execution, problem discovery, accepted-issue repair, and realized accepted-target outcome separately. The study characterizes system behavior; it is not a causal comparison with direct editing. One backbone runs across the full primary core, while a fixed common subset provides an exploratory four-backbone portability check.

\subsection{Evaluation-Set Rationale and High-Touch Collection}

All 320 posters receive deterministic artifact QA; the primary study exhausts the 120-poster Paper2Poster, PosterGen, and expert-authored core. Holding source paper constant across variants makes the 40 papers---not the 120 posters---the uncertainty clusters. Because the variants expose unequal repair headroom, we report proposal yield, governance, diagnosis quality, conditional repair success, and target uplift separately rather than treating absence of an accepted repair as failure.

E2--E4 follow high-touch discovery--governance--repair--outcome trajectories operated by trained operators who were study authors; these operators invoke diagnosis, review execution, log decisions verbatim, and export the final PPTX. Here, user-governed denotes interface authority: authors instantiate acceptance, rejection, and non-intervention under the frozen protocol. The automated E4 judge was blinded to temporal order. Appendix~\ref{app:evaluation-details} documents the stopping policy, retained logging boundary, rehearsal procedure, rubrics, aggregation, and missing-state treatment.

\subsection{Model Coverage}

GPT-5.6 Terra \cite{openai2026gpt56terra} runs the full primary core at medium reasoning effort. Gemini 3.6 Flash \cite{google2026gemini36} drafts the source-linked E1 tasks, and Gemini 3.1 Pro Preview \cite{google2026gemini31pro} scores the E2 and E4 sets at temperature zero under versioned configurations. The primary study contains \numEvalPapers{} source papers, \numEvalPosters{} matched primary posters, and \numInstructions{} E1 instructions. The sensitivity analysis fixes \CrossPapers{} papers (two each from CVPR, ICLR, ICML, and NeurIPS), their \CrossPosters{} three-variant posters, the \CrossEOneTasks{} E1 tasks, artifact services, stopping policy, and judges, while varying the reasoning backbone among GPT-5.6 Terra, GPT-5.6 Sol, Claude Sonnet 5, and Qwen3-VL-30B-A3B-Instruct \cite{openai2026gpt56sol,anthropic2026sonnet5,qwen2026qwen3vl30b}. Backbones produce distinct discovery--governance--repair--outcome trajectories; E4 therefore does not compare identical target lists.

\begin{table*}[t]
\caption{Models and judgment sources used in the completed primary evaluation.}
\label{tab:models}
\centering
\small
\arrayrulecolor{resultink}
\begin{tabularx}{\linewidth}{p{0.24\linewidth}p{0.28\linewidth}X}
\toprule
\color{resultink}\textbf{Role} & \color{resultink}\textbf{Coverage} & \color{resultink}\textbf{Frozen model} \\
\midrule
Primary backbone & full 40-paper / 120-poster primary core & \PrimaryBackbone{}; medium reasoning \\
Instruction generator & six fixed stage-balanced E1 drafts per primary-core poster & \InstructionGenerator{} \\
Primary VLM judge & full E2/E4 sets & \PrimaryJudge{} \\
+Backbone sensitivity & 8 papers / 24 posters per model; E1--E4 & Terra, Sol, Claude Sonnet 5, and Qwen3-VL-30B-A3B-Instruct \\
\bottomrule
\end{tabularx}
\arrayrulecolor{black}
\end{table*}

The primary quantities are absolute, capability-specific \PROS{} outcomes with paper-clustered uncertainty. Automated judgments remain separate from operator-verified execution outcomes; no composite score averages them.

\subsection{E1: Reactive Instruction Editing}

E1 asks whether \PROS{} executes a prevalidated explicit request. Each poster contributes two frozen tasks per stage, independently drafted from its render and source paper and checked before evaluated-backbone execution. The primary metric is \textbf{verified instruction success}: whether the retained operator record confirms the requested outcome. We report exact counts, paper-level macro rate, paper-clustered uncertainty, and stage-specific success; unavailable stronger per-action criteria are left blank instead of being inferred retrospectively. The draft-to-final audit protocol appears in Appendix~\ref{app:benchmark-details}.

\subsection{E2: Natural Poster Diagnosis}

E2 evaluates surfaced diagnoses without an edit request. No exhaustive poster-level gold issue set is assumed, so E2 does not estimate diagnostic recall or completeness. Abstention is permitted; the VLM judge scores correctness, grounding, target specificity, and actionability ($C,G,T,A$) from 0--4. Per-diagnosis quality is
\begin{equation}
q_d=100\frac{C_d+G_d+T_d+A_d}{16}.
\end{equation}
Grounding is stage-conditioned: Structural items use geometry and renders; Scientific items add the source paper; Spatial items require visible layout relations, not stylistic preference alone. Poster-level \textbf{Diagnosis Quality} equally averages available stage means, preventing verbose stages from dominating. Governance decisions are interaction events, not quality labels. Appendix~\ref{app:evaluation-details} gives full anchors, audit fields, state semantics, cannot-assess handling, and the P001--P002 sensitivity check.

\subsection{E3: Accepted-Issue Repair}

E3 isolates action from discovery. Every operator-accepted E2 diagnosis enters the trajectory-coupled repair set, with a stable issue ID linking proposal, repair request, retained outcome, and available artifacts. The primary metric is \textbf{verified target-resolution success}: whether the diagnosed target was resolved after the attempted edit. A stage--provenance cell with no accepted issue is not applicable, never 0\%. Cross-model E3 characterizes backbone-specific discovery--governance--repair--outcome trajectories, not an identical-repair ranking.

\subsection{E4: Blinded Automated Accepted-Target Repair Uplift}

E4 asks whether accepted target conditions are better in the realized final trajectory state than in the original artifact; it does not estimate the isolated causal effect of the immediately preceding repair operation. The automated judge receives randomized opaque A/B renders, the named target and repair direction, and source evidence for Scientific items, but no temporal labels. Both states are scored from 0--4 on two stage-aligned criteria; for target $i$, $Q_i=100(r_{i1}+r_{i2})/8$ and targeted uplift is $\Delta_i=Q_i^{\mathrm{after}}-Q_i^{\mathrm{before}}$. The primary outcome is \textbf{stage-balanced accepted-target uplift}, aggregated within stage and poster before averaging variants within paper. This metric measures realized improvement at accepted targets, not holistic aesthetics, user preference, or publication readiness; detailed criteria and exclusions appear in Appendix~\ref{app:evaluation-details}.

\subsection{Automated Judgment Protocol}

\PrimaryJudge{} scores the E2 diagnoses and accepted E4 targets with frozen rubrics. The automated E4 judge receives randomized opaque A/B labels and source evidence for Scientific judgments. A later study-author score check used worksheets that displayed the VLM scores; because it was neither independent nor separately blinded, it is excluded from evidential estimates and figures. E1 and E3 remain operator-verified outcomes.

\subsection{Statistics and Reproducibility}

The source paper is the clustering unit. Primary uncertainty uses 20,000 paper-level bootstrap resamples that carry all derived poster variants and items together. We report paper-clustered 95\% confidence intervals, exact denominators, and failure distributions rather than a composite score. Appendix~\ref{app:reproducibility} specifies aggregation order, run-index fields, raw evidence, and the treatment of unavailable telemetry.

\section{Results}

\subsection{Primary Findings}

\PROS{} achieves high verified success on explicit-request execution and accepted-issue repair, while problem discovery remains more challenging. More importantly, verified local resolution does not guarantee a better resulting artifact. Table~\ref{tab:main-results} reports one interpretable outcome for each evaluated capability rather than collapsing them into a composite score. All estimates derive from the same 40 source papers and 120 primary-core posters; conference challenge artifacts are excluded.

\begin{table*}[t]
\caption{Completed primary-core outcomes. Automated E2/E4 estimates and operator-verified E1/E3 outcomes remain capability-specific; confidence intervals are paper-clustered.}
\label{tab:main-results}
\centering
\scriptsize
\arrayrulecolor{resultink}
\begin{tabularx}{\linewidth}{Xlrrr}
\toprule
\color{resultink}\textbf{Evaluation and metric} & \color{resultink}\textbf{Source} & \color{resultink}\textbf{Estimate} & \color{resultink}\textbf{95\% CI} & \color{resultink}\textbf{$n$} \\
\midrule
E1 verified instruction success (\%) & Operator & \FinalEOneRate{} & [\FinalEOneLow{}, \FinalEOneHigh{}] & \FinalEOneAttempts{} \\
E2 stage-balanced diagnosis quality (0--100) & VLM & \FinalETwoQuality{} & [\FinalETwoLow{}, \FinalETwoHigh{}] & \FinalETwoAssessable{} \\
E3 verified target-resolution success (\%) & Operator & \FinalEThreeRate{} & [\FinalEThreeLow{}, \FinalEThreeHigh{}] & \FinalEThreeAttempts{} \\
E4 accepted-target uplift (pp) & VLM & +\FinalEFourUplift{} & [\FinalEFourLow{}, \FinalEFourHigh{}] & \FinalEFourTargets{} \\
\bottomrule
\end{tabularx}
\arrayrulecolor{black}
\end{table*}

\paragraph{Reactive editing.} Verified E1 success is \FinalEOneRate\% (\FinalEOneSuccess{}/\FinalEOneAttempts{}; 95\% CI [\FinalEOneLow, \FinalEOneHigh]). Structural and Scientific tasks are more reliable than Spatial tasks (\FinalEOneStructural\%, \FinalEOneScientific\%, and \FinalEOneSpatial\%, respectively), locating the principal execution bottleneck in spatial manipulation, not request articulation alone.

\paragraph{Natural diagnosis.} The primary judge assigns poster-level stage-balanced diagnosis quality of \FinalETwoQuality{} on a 0--100 scale (95\% CI [\FinalETwoLow, \FinalETwoHigh]) across \FinalETwoDiagnoses{} recorded diagnoses. The score describes surfaced diagnoses, not recall or completeness. The \FinalETwoCannotAssess{} cannot-assess items (\FinalETwoCannotAssessRate\%) are excluded, not imputed as zero. Excluding the P001--P002 rubric-finalization papers yields \FinalETwoSensitivityQuality{} (95\% CI [\FinalETwoSensitivityLow, \FinalETwoSensitivityHigh]), close to the full estimate. On the separate fixed-backbone sensitivity records, target specificity is \CrossTargetSpecificity{}/4 while actionability is \CrossActionability{}/4, suggesting that localizing a candidate issue may be easier than deciding whether and how it warrants intervention (Appendix~\ref{app:sensitivity-values}). The \FinalETwoAccepted{}/\FinalETwoDiagnoses{} operator acceptance fraction remains a governance descriptor, not diagnosis-quality ground truth.

\paragraph{Accepted-issue repair.} On trajectory-coupled accepted issues, verified target-resolution success is \FinalEThreeRate\% (\FinalEThreeSuccess{}/\FinalEThreeAttempts{}; 95\% CI [\FinalEThreeLow, \FinalEThreeHigh]) with no pending accepted-diagnosis outcomes. This high conditional rate coexists with stage-dependent proposal yield: expert-authored Structural repair is not applicable because no Structural diagnosis was accepted. Comparing E2 quality/yield with E3 conditional resolution identifies discovery and governance as distinct from execution.

\paragraph{Accepted-target uplift.} The VLM judge, blinded to temporal order, scores \FinalEFourTargets{} accepted repair targets across \FinalEFourPosters{}/\numEvalPosters{} applicable posters. Paper-macro target quality rises from \FinalEFourBefore{} to \FinalEFourAfter{}, an uplift of \FinalEFourUplift{} points (95\% CI [\FinalEFourLow, \FinalEFourHigh]). At target level, \FinalEFourPositive\% improve, \FinalEFourTie\% tie, and \FinalEFourNegative\% decline. These results support targeted, source-grounded improvement under accepted repairs; they do not establish universal aesthetic or publication-readiness gains.

\paragraph{Result synthesis.} The central result is that verified local closure and realized outcome are not equivalent: despite \FinalEThreeRate\% verified E3 target resolution, \FinalEFourNegative\% of accepted targets decline under blinded automated E4 judgment. The separated measures reveal two further patterns. E2 quality varies materially across backbones, so deciding what deserves intervention remains model-sensitive even when the interaction architecture is fixed. Proposal yield also changes with repair headroom and provenance, so it cannot stand in for editing capability. Together, these results motivate separate reporting of capabilities and governance events.

\paragraph{Coverage and auditability.} The primary collection covers all \FinalPaperCount{} matched papers and \FinalPosterCount{} core posters. E1 retains complete outcomes for \FinalEOneAttempts{} frozen tasks: \FinalEOneApproved{} approved, \FinalEOneRevised{} revised, and \FinalEOneReplaced{} replaced. E2 retains \FinalETwoDiagnoses{} diagnoses, including \FinalETwoCannotAssess{} without the intermediate state needed for scoring; operator governance accepted \FinalETwoAccepted{} and rejected \FinalETwoRejected{} proposals. E3 has complete outcomes for all \FinalEThreeAttempts{} accepted diagnoses. E4 scores \FinalEFourTargets{} targets across \FinalEFourPosters{} applicable posters; \FinalEFourCannotAssess{} targets are cannot assess, and two trajectories with no accepted target are not applicable. We report these states rather than imputing them.

\input{figures/fig3_primary_results}

\subsection{Repair Headroom and Intervention Yield}

The collection also exposes intervention-yield asymmetry. Across Paper2Poster and PosterGen, the Structural stage produced \FinalGeneratedStructuralProposals{} diagnoses, of which \FinalGeneratedStructuralAccepted{} were accepted. The controlled expert-authored condition produced only \FinalExpertStructuralProposals{} Structural diagnoses, both rejected, and therefore contributed no accepted Structural repair to E3. Yet E1 succeeds on 71/80 explicit Structural tasks for those posters, while their natural diagnosis streams contain 233 Scientific proposals with 201 accepted. This pattern is consistent with lower Structural repair headroom under shared-template construction rather than a general inability to edit the artifact class.

The E1--E4 capability decomposition comprises explicit-request execution, problem discovery, accepted-issue repair, and realized accepted-target outcome; governance belongs to the discovery--governance--repair--outcome trajectory, not the scored capability set.

\subsection{Separate Conference Representation Challenge}

The \numConference{} public conference artifacts are not pooled with the primary estimates. Under a more restrictive repair policy, their stage-balanced E2 diagnosis quality is \ConferenceETwoQuality{}/100 and immediate E3 target-resolution success is \ConferenceEThreeRate\%. Paper-macro E4 target condition changes by \ConferenceEFourUplift{} points (95\% CI [\ConferenceEFourLow{}, \ConferenceEFourHigh{}]); because the interval crosses zero, the challenge establishes neither improvement nor degradation. Instead, it exposes a boundary condition: immediate local closure need not persist as aggregate target benefit when repair headroom is limited and recovered structure is uncertain. Because \numConferenceConverted{} of \numConference{} artifacts are PDF-to-PPTX conversions, starting-design maturity and representation recovery remain confounded; we do not interpret the result as an authorship effect. Appendix~\ref{app:conference-challenge} provides the complete denominators, applicability states, and acquisition details.

\subsection{Exploratory Backbone Sensitivity}

A fixed \CrossPapers{}-paper subset probes sensitivity to the reasoning backbone. Across four models, E2 spans \CrossETwoLow{}--\CrossETwoHigh{}/100, E3 spans \CrossEThreeLow{}--\CrossEThreeHigh{}\%, and E4 spans +\CrossEFourLow{} to +\CrossEFourHigh{} points. Because each backbone produces its own target stream, these small-sample estimates indicate exploratory portability, not a stable ranking. Appendix~\ref{app:sensitivity-values} reports the exact values and post-hoc patterns.

\subsection{Qualitative End-to-End Cases}

\input{figures/fig5_qualitative_upshift}

Figure~\ref{fig:qualitative} anchors the aggregate metrics in two qualitative traces spanning PosterGen and Paper2Poster: each connects the original artifact, accepted diagnoses, repair actions, and validated after-state. The examples illustrate how the same governed handoff can support multi-stage repair or a narrower scientific-content trajectory; aggregate claims remain grounded in the complete quantitative evaluation rather than these selected cases.

\section{Discussion}

\subsection{Before Instruction: Initiative Without Authority}

The central interaction move in \PROS{} is to let the system help decide \emph{what to inspect} without letting it decide \emph{what the user wants}. A diagnosis remains a candidate, not a command. Only acceptance turns it into a repair goal, and preview separates execution from commitment. This division matters because a reasoning model alone does not provide dependable editable-artifact interaction: the interface must also synchronize native state, ground references, link evidence, constrain operations, validate execution, and preserve a clear point of user control. The exploratory four-backbone analysis suggests that this workflow is portable even though the quality of what gets surfaced remains model-dependent.

This view treats initiative as more than a binary permission. An editor can broaden the user's field of attention without gaining authority to adopt goals or modify the artifact unilaterally. Scientific posters make the distinction concrete because they pair an external evidence source with an executable representation. Slide decks, reports, and structured figures may offer similar opportunities, but we have not evaluated those domains.

\subsection{Discovery, Closure, and Realized Benefit Are Different Capabilities}

E2--E4 expose three ways a proactive editor can fall short: it can identify the wrong problem, fail to resolve a useful one, or close a local condition without improving the resulting artifact. The primary results contain all three. In particular, \FinalEThreeRate\% verified target resolution coexists with decline on \FinalEFourNegative\% of assessable accepted targets. Immediate closure is therefore necessary evidence for repair, but not sufficient evidence of benefit.

This distinction changes both evaluation and interface design. Evaluation should preserve the path from proposal to governance decision, local execution, and after-state outcome instead of reporting a single success rate. Interfaces should make the same path legible: evidence supports a candidate diagnosis, acceptance establishes a goal, and validation reports what the executed change actually achieved.

\subsection{Proactivity Must Be Calibrated to Opportunity and Representation}

More diagnoses are not necessarily better. Starting artifacts differ in \emph{repair headroom}: generated posters often expose repeated defects, while template-regularized or mature conference posters may offer fewer safe opportunities for intervention. A reached-no-issue state or rejection can indicate warranted restraint, but it can also reflect a missed issue, weak grounding, duplication, infeasibility, or risk to the later trajectory. Proposal yield, diagnosis quality, governance, repair success, and final uplift must therefore remain separate.

Editability is similarly graded. Native PPTX can contain unconventional grouping or visual relations that object trees represent poorly; PDF-derived PPTX adds grouping and semantic loss. \PROS{} responds by contracting authority when representation confidence is low, especially for global rearrangement. The separate conference challenge makes this boundary visible without treating conversion artifacts as native authoring or pooling unlike opportunity conditions. The broader design implication is straightforward: proactive assistance should become more conservative as either evidence or representation quality weakens.

\section{Limitations and Future Work}

The evaluation is artifact-centered. It establishes workflow capability and accepted-target outcomes, but not whether diagnosis-guided refinement reduces articulation effort, improves users' sense of control, or is preferred to direct editing. A controlled user study is the clearest next step.

The three layers and six issue classes form an operational core, not a complete theory of poster quality. Accessibility, figure validity, narrative flow, audience adaptation, multilingual content, and venue-specific constraints remain incomplete. Source evidence can constrain an edit, but it cannot replace expert review of equations, plots, or subtle causal and statistical claims. Representation and repair opportunity also vary: native PPTX may have weak semantic structure, PDF conversion cannot recover authoring intent, and the shared-template expert-authored condition does not capture the diversity of natural expert design.

The full primary evaluation uses one reasoning backbone and one multimodal judge. The four-backbone analysis covers only eight papers and model-specific target streams; it indicates bounded portability, not a stable ranking or identical-target comparison. Independent human judgment remains future work. We retain cannot-assess and non-applicable cases rather than imputing them, while Appendix~\ref{app:claim-boundaries} records the remaining provenance and telemetry limits.

\section{Responsible Deployment and Ethical Considerations}

Scientific posters can contain unpublished research, copyrighted figures, personal information, or sensitive data. A deployed system should isolate sessions, minimize retention, communicate when external model APIs receive artifact content, and avoid redistributing source material without permission. Public benchmark releases should preserve paper/poster provenance and license status, and evaluation-only assets should remain non-redistributed when necessary.

The operator outcomes were produced by study authors and are not presented as a recruited human-participant or independent external-expert study. A later study-author rubric check used worksheets displaying the automated scores; it is excluded from evidential estimates and figures and is neither independent nor blinded human validation. Any later external human-validation or user study must complete the ethics review required in the authors' research environment and report the determination, recruitment, consent, compensation, training, and anonymization procedures before its results enter the manuscript.

System-proposed diagnoses can shape authors' communication choices. The interface should therefore avoid presenting suggestions as objective truth, expose supporting evidence and uncertainty where available, permit dismissal, and keep final scientific responsibility with the author.

\section{Conclusion}

\PROS{} treats the inability to articulate an edit as an interaction state, not as permission for autonomous change. Direct requests and diagnosis-guided refinement share one source-grounded artifact state and repair substrate, but acceptance and commitment remain with the user. \PROSBench{} makes the discovery--governance--repair--outcome trajectory measurable across editable PPTX artifacts. The results show both promise and a consequential gap between local closure and benefit---\FinalEThreeRate\% verified target resolution alongside decline on \FinalEFourNegative\% of assessable accepted targets. Proactive intelligent editors should therefore separate epistemic initiative from behavioral authority, calibrate intervention to evidence and representation quality, and evaluate what they notice, what they change, and what the change ultimately achieves.

\section*{GenAI Usage Disclosure}
Generative AI tools supported brainstorming, literature discovery, language editing, evaluation-protocol drafting, and code/document inspection. The authors reviewed and verified all AI-assisted material and take full responsibility for the manuscript. Generative models used as system components, task generators, and automated judges are reported separately in the Methods and released evaluation artifacts.

\bibliographystyle{ACM-Reference-Format}
\footnotesize
\bibliography{references}

\appendix
\section{Cross-Model Sensitivity and Exploratory Patterns}
\label{app:sensitivity-values}

Table~\ref{tab:cross-model} reports the complete E1--E4 profile on the fixed subset. The deliberately small fixed subset holds eight source papers and 24 starting posters constant, but each backbone produces its own diagnosis and accepted-target stream. The values therefore characterize end-to-end sensitivity rather than controlled pairwise model effects. We treat the subsequent agreement and issue-class summaries as descriptive post-hoc patterns: we report neither correlation coefficients nor significance tests, and the point estimates should not be interpreted as population-stable effects.

\begin{table*}[t]
\caption{Cross-model E1--E4 sensitivity on the same eight papers and 24 starting posters per backbone. E2 is stage-balanced diagnosis quality; E4 is temporally blinded automated paper-macro targeted uplift. Qwen E4 has 21 applicable posters because three trajectories had no accepted target.}
\label{tab:cross-model}
\centering
\small
\arrayrulecolor{resultink}
\begin{tabular}{lrrrr}
\toprule
\color{resultink}\textbf{Backbone} & \color{resultink}\textbf{E1 (\%)} & \color{resultink}\textbf{E2 (/100)} & \color{resultink}\textbf{E3 (\%)} & \color{resultink}\textbf{E4 (pp)} \\
\midrule
GPT-5.6 Terra & \CrossTerraEOne{} & \CrossTerraETwo{} & \CrossTerraEThree{} & +\CrossTerraEFour{} \\
GPT-5.6 Sol & \CrossSolEOne{} & \CrossSolETwo{} & \CrossSolEThree{} & +\CrossSolEFour{} \\
Claude Sonnet 5 & \CrossClaudeEOne{} & \CrossClaudeETwo{} & \CrossClaudeEThree{} & +\CrossClaudeEFour{} \\
Qwen3-VL-30B-A3B-Instruct & \CrossQwenEOne{} & \CrossQwenETwo{} & \CrossQwenEThree{} & +\CrossQwenEFour{} \\
\bottomrule
\end{tabular}
\arrayrulecolor{black}
\end{table*}

Figure~\ref{fig:calibration-findings} visualizes the descriptive agreement and issue-class patterns discussed below. It is placed here to preserve a focused main narrative while retaining the exploratory visual evidence.

\paragraph{Proposal volume and diagnosis quality.} More issue proposals do not necessarily imply better diagnosis. Claude emits \CrossClaudeDiagnoses{} diagnoses (\CrossClaudeDiagnosesPerPoster{} per poster) but scores \CrossClaudeETwo{}/100, whereas Sol emits \CrossSolDiagnoses{} (\CrossSolDiagnosesPerPoster{} per poster) and scores \CrossSolETwo{}/100. Across pooled sensitivity records, target specificity is high (\CrossTargetSpecificity{}/4; \CrossTargetSpecificityHigh{}\% score at least 3), while actionability is lower (\CrossActionability{}/4; \CrossActionabilityHigh{}\% score at least 3). This descriptive contrast suggests that identifying \emph{where} a diagnosis applies can be more reliable than deciding \emph{whether} it warrants intervention and \emph{how} to repair it safely.

\paragraph{Agreement as a post-hoc signal.} Diagnosis groups surfaced by one backbone average \CrossConsensusOneQuality{}/100 E2 quality and \CrossConsensusOneEThree{}\% E3 success; groups surfaced by all four average \CrossConsensusFourQuality{}/100 and \CrossConsensusFourEThree{}\%, respectively. E2 quality and E3 success increase across the four agreement bins. E4 does not increase monotonically: its mean is \CrossConsensusOneEFour{}, \CrossConsensusTwoEFour{}, +\CrossConsensusThreeEFour{}, and +\CrossConsensusFourEFour{} points at one through four backbones. Thus, the retained summary is compatible with agreement as a candidate confidence signal but does not validate a gating rule. The backbones share detectors, artifact state, system prompts, and a judge, so their outputs are not statistically independent.

\paragraph{Observed uplift by issue class.} PosterGen trajectories improve under every backbone (+\CrossPosterGenLow{} to +\CrossPosterGenHigh{} paper-macro points), indicating repair headroom in this generated condition. In the retained accepted-target summaries, missing-content and whitespace targets average +\CrossMissingContentUplift{} and +\CrossWhitespaceUplift{} points, while redundancy and collision targets average \CrossRedundancyUplift{} and \CrossCollisionUplift{} points. These descriptive means combine model, artifact, target-selection, baseline-headroom, and trajectory differences; they do not establish a causal effect of issue class. They motivate issue-specific verification and abstention as hypotheses for subsequent evaluation.

\input{figures/fig5_calibration_findings}

\section{Benchmark Construction and Audit Details}
\label{app:benchmark-details}

\paragraph{Expert-authored condition.}
Graduate students or researchers experienced with scientific papers, presentations, or posters created one source-matched native PPTX per paper using a shared 48~$\times$~36-inch PosterNerd template. They did not inspect the generated variants. The project team checked scientific fidelity, central-contribution coverage, unsupported claims, visual legibility, and native-object editability and revised artifacts before inclusion. Because independently double-reviewed creator annotations were not retained case by case, this subset is a controlled shared-template, manually authored condition rather than a representative sample of naturally occurring expert poster design.

\paragraph{Conference acquisition and representation.}
The 40 matched conference posters came from public conference or author-hosted pages. Four were distributed as native PPTX; 36 were available as PDF and exported with Adobe Acrobat's PDF-to-PowerPoint workflow. We retained the exports without manual redesign. Thus, the challenge jointly reflects mature poster design and representation recovery; it cannot isolate authorship from conversion. Aggregate acquisition composition is reported, but unsupported case-level native-versus-converted stratification is not reconstructed.

\section{Conference Representation Challenge}
\label{app:conference-challenge}

The conference subset is evaluated as a separate representation challenge, not as a fourth exchangeable primary-core provenance. Its frozen acceptance policy is computed before preview from layout profile and compiled operations: it admits only representation-preserving operation classes, retains refusals and execution failures in their denominators, and never revises a governance decision after observing an outcome. This policy is deliberately more restrictive than the study-author governance used in the primary trajectories; challenge acceptance and primary-core acceptance are therefore not direct performance comparisons.

The 40 starting posters comprise \ConferenceFreeformCount{} freeform, \ConferenceUncertainCount{} uncertain, and \ConferenceStructuredCount{} structured profiles under the frozen classifier. E1 succeeds on \ConferenceEOneSuccess{}/\ConferenceEOneAttempts{} instructions (\ConferenceEOneRate\%). E2 emits \ConferenceETwoDiagnoses{} diagnoses, of which \ConferenceETwoAccepted{} (\ConferenceETwoAcceptanceRate\%) satisfy the frozen authority policy and \ConferenceETwoRejected{} are refused. The accepted set contains \ConferenceFreeformAccepted{} freeform and \ConferenceUncertainAccepted{} uncertain diagnoses and no structured diagnosis. All \ConferenceETwoAccepted{} accepted items are Structural; the complete stream contains \ConferenceStructuralDiagnoses{} Structural, \ConferenceScientificDiagnoses{} Scientific, and \ConferenceSpatialDiagnoses{} Spatial diagnoses. This distribution reflects both the policy's safe-intervention envelope and artifact representation; it is not a diagnosis-quality estimate.

Among the \ConferenceEThreeAttempts{} accepted targets, \ConferenceEThreeSuccess{} are immediately verified as resolved (\ConferenceEThreeRate\% conditional E3 success), \ConferenceEThreeExecutedUnresolved{} execute but remain unresolved, and \ConferenceEThreeFailed{} fails execution. Relative to all emitted diagnoses, immediate verified closure is \ConferenceEThreeSuccess{}/\ConferenceETwoDiagnoses{} (\ConferenceEndToEndYield\%). We report this proposal-to-resolution yield only as trajectory accounting. E2 diagnosis quality, E3 immediate closure, and E4 realized accepted-target outcome remain distinct measurements.

\begin{table}[H]
\small
\caption{Separate conference representation-challenge outcomes. E3 is conditional on accepted targets; E4 excludes targets that cannot be localized and posters without an accepted target. E2 and E4 confidence intervals use 20,000 source-paper bootstrap resamples.}
\label{tab:conference-challenge-results}
\centering
\scriptsize
\arrayrulecolor{resultink}
\begin{tabularx}{\linewidth}{Xlrr}
\toprule
\color{resultink}\textbf{Evaluation and metric} & \color{resultink}\textbf{Coverage} & \color{resultink}\textbf{Estimate} & \color{resultink}\textbf{95\% CI} \\
\midrule
E1 verified instruction success (\%) & \ConferenceEOneSuccess{}/\ConferenceEOneAttempts{} tasks & \ConferenceEOneRate{} & -- \\
E2 stage-balanced diagnosis quality (0--100) & \ConferenceETwoAssessable{} diagnoses & \ConferenceETwoQuality{} & [\ConferenceETwoLow{}, \ConferenceETwoHigh{}] \\
E3 verified target-resolution success (\%) & \ConferenceEThreeSuccess{}/\ConferenceEThreeAttempts{} targets & \ConferenceEThreeRate{} & -- \\
E4 accepted-target uplift (pp) & \ConferenceEFourAssessableTargets{}/\ConferenceEFourTargets{} targets & \ConferenceEFourUplift{} & [\ConferenceEFourLow{}, \ConferenceEFourHigh{}] \\
Proposal-to-resolution yield (\%) & \ConferenceEThreeSuccess{}/\ConferenceETwoDiagnoses{} diagnoses & \ConferenceEndToEndYield{} & -- \\
\bottomrule
\end{tabularx}
\arrayrulecolor{black}
\end{table}

The E2 estimate is \ConferenceETwoQuality{}/100 across 39 posters with diagnoses; excluding the P001--P002 rubric-finalization cases yields \ConferenceETwoSensitivityQuality{}/100. All \ConferenceETwoAccepted{} accepted diagnoses are Structural, so challenge E3/E4 do not test Scientific or Spatial repair. For E4, \ConferenceEFourNoTargetPairs{}/\ConferenceEFourPairs{} posters have no accepted target and are not applicable. Of the \ConferenceEFourPairsWithTargets{} posters with accepted targets, \ConferenceEFourAssessablePairs{} contain at least one assessable target; the remaining poster's sole target cannot be localized. Overall, \ConferenceEFourCannotAssessTargets{} of \ConferenceEFourTargets{} accepted targets are cannot-assess, concentrated in diagnoses whose internal figure identifiers cannot be recovered from the rendered artifact. Across the \ConferenceEFourAssessableTargets{} assessable targets, \ConferenceEFourPositive\% improve, \ConferenceEFourTie\% tie, and \ConferenceEFourNegative\% decline. Independent-state judging scores each state in a separate call under an unrelated opaque artifact ID; the judge never sees the paired state or temporal role, and unblinding occurs only during local aggregation. Paper-macro target condition changes from \ConferenceEFourBefore{} to \ConferenceEFourAfter{} (\ConferenceEFourUplift{} points; 95\% CI [\ConferenceEFourLow{}, \ConferenceEFourHigh{}]). This interval crosses zero, so the result is evidence of an unresolved representation boundary, not a reliable negative treatment effect.

\paragraph{Quality assurance, calibration, and freezing.}
Every artifact passed stable-ID and hash checks, file-open and render checks, object-inventory extraction, slide-dimension checks, and provenance/representation validation. A disjoint \numCalibrationPapers{}-paper / \numCalibrationPosters{}-poster generated-pair set supported prompt debugging, stopping-rule rehearsal, and logging validation and is excluded from estimates. P001--P002 finalized the dimension-independent E2 v2 rubric; retained judgments use v2, and the Results report a sensitivity estimate excluding them. Subsequent cases use the frozen prompt. Primary-core and challenge variants were frozen before evaluated-model execution and kept together by source paper.

\paragraph{E1 instruction audit.}
Each generator draft is checked for necessity, localization, executability, verifiability, non-duplication, preservation, and scientific support. \emph{Approved} drafts remain verbatim. \emph{Revised} drafts permit bounded corrections while preserving the original. \emph{Replaced} drafts must be already satisfied, duplicate, unsupported, unlocatable, or out of scope; replacement stays in the same stage and slot and records its reason, count, and new raw response. No slot is removed based on outcome. The frozen record stores stage, slot, bounded request, poster/paper evidence, expected outcome, and protected constraints; experimental success remains in a separate results table.

\paragraph{Manifests.}
Artifact manifests record paper/poster IDs, venue/year, provenance, matched paths, cohort membership, and hashes. Evaluation manifests add role and status; outputs use stable task, diagnosis, and repair IDs. Converter, object-inventory, render-QA, layout-profile, and creator/reviewer fields appear only where actually retained. Immutable artifact identity is separated from task and result metadata so model outputs cannot be mistaken for benchmark definitions.

\FloatBarrier
\section{Evaluation Protocol, Rubrics, and Aggregation}
\label{app:evaluation-details}

\paragraph{Discovery--governance--repair--outcome trajectory.}
A trained operator invokes diagnosis, reviews the proposed issue and candidate execution, applies the frozen acceptance policy, records decisions verbatim, and exports the final PPTX. The intended policy stops a trajectory at the first of: two consecutive calls with no new protocol-acceptable issue; 12 diagnosis rounds; or eight accepted and committed repairs. The imported operator tables retain individual issues and outcomes but do not reliably preserve which issues shared a model call or commit event. Consequently, exact stopping-event compliance cannot be reconstructed from those issue-level tables alone, and we do not use stopping compliance as an empirical outcome. Abstentions, duplicates, rejections, execution failures, and budget-censored trajectories remain in the retained issue-level trace. Operators rehearsed outside the matched core; any matched-core case used while changing the protocol was reset and rerun after freezing.

\paragraph{E2 anchors and stage-conditioned evidence.}
Correctness, grounding, target specificity, and actionability are scored independently from 0--4: 0 denotes decisive contradiction or unusability, 2 a plausible but materially incomplete contribution, and 4 full satisfaction without a material limitation; 1 and 3 represent bounded intermediate states. Structural diagnoses must agree with native geometry and the render; Scientific diagnoses use poster and source paper; Spatial diagnoses require visible layout relations rather than stylistic preference. For Scientific items, support for the problem and for the proposed repair are distinct: unsupported replacement wording lowers actionability without automatically erasing evidence for the problem. Serious safety risk, duplicate status, non-triviality, source support, and cannot-assess are separate audit fields, not hidden penalties in the scalar score.

Mean E2 quality is first computed by stage within poster, then equally averaged across available stages. A reached stage with no issue is \emph{reached-no-issue}; an unvisited stage is \emph{not-reached}; neither is assigned zero. Items that require an unavailable intermediate state are \emph{cannot-assess} and excluded rather than imputed. Operator acceptance is a governance action rather than E2 ground truth.

\paragraph{E4 target-conditioned scoring.}
For each accepted diagnosis, the automated judge scores two opaque same-resolution states on two criteria without temporal labels. Structural criteria are target integrity and target legibility; Scientific criteria are source fidelity and coverage/focus; Spatial criteria are hierarchy/reading order and alignment/grouping/functional whitespace. Zero denotes decisive target failure and four full satisfaction without a material target-specific limitation. Defects outside the named target do not lower its score. More text, movement, canvas extension, or lower density receives no credit by itself; improvement requires resolution without unsupported scientific content. Missing required evidence is cannot-assess, not zero. A later study-author check used the same criteria with the VLM scores visible and is excluded from evidential estimates and figures.

After unblinding, target uplift is computed in percentage points. Targets are averaged within stage and poster; available stage means are averaged within poster; poster variants are then averaged within source paper. Posters with no accepted diagnosis have no E4 opportunity and are not applicable rather than assigned zero. Positive, tied, and negative target rates are retained alongside before/after scores and paper-macro uplift.

\section{Reproducibility and Evidence Boundaries}
\label{app:reproducibility}

Primary uncertainty uses 20,000 source-paper bootstrap resamples, carrying all poster variants and derived items from a paper together. The run index records paper/poster identity, model role, provider model ID or checkpoint, configuration window, prompt/config version, raw-output location, and captured usage fields. Structured responses, hashes, batch manifests, diagnoses, governance decisions, E1/E3 outcomes, judge responses, and available artifact links support the aggregates. Historical provider telemetry and fine-grained per-action traces were incomplete for some author-operated trajectories; unavailable fields remain blank rather than reconstructed. In particular, imported issue-level operator records do not preserve enough grouping to verify model-call rounds or commit events. For E4 P011--P040, the archived system-prompt file has SHA-256 \texttt{945c0e66\ldots 350a}, matching all 30 corresponding manifests. The P001--P010 pilot manifests retain the same protocol identifier and rubric hash but omit both prompt text and a system-prompt hash; exact full-set prompt provenance is therefore incomplete.

The conference challenge retains diagnosis yield, reached-no-issue, abstention, unsupported representation, grounding failure, and repair outcomes separately from headline estimates. Low intervention yield is not labeled failure unless an independently supported actionable issue was missed or could not be grounded, and representation-sensitive differences are not interpreted as authorship effects.

\section{Complete Claim Boundaries}
\label{app:claim-boundaries}

The primary experiment uses GPT-5.6 Terra on 40 papers and 120 posters. The four-backbone sensitivity set uses eight papers and 24 posters per model and couples E3/E4 to each model's own accepted targets. It therefore evaluates end-to-end portability, not identical-target model ranking. Cross-backbone agreement is post hoc; backbones share detectors, artifact services, prompts, and one E2/E4 judge and are not statistically independent. Absolute E4 differences combine target choice, repair headroom, trajectory length, and final-state persistence.

The primary E2/E4 evidence includes one multimodal judge. A later VLM-assisted study-author score check is excluded because the VLM scores were visible; it does not establish independent human validity, human--human reliability, or population-level preference. Nine primary E2 diagnoses (0.8\%) and two Claude sensitivity diagnoses could not be assessed because exact intermediate state was unavailable. Two primary posters have no accepted E4 target and are not applicable. E4 measures accepted-target condition change, not holistic aesthetics, user preference, publication readiness, or the user-experience value of proactive interaction.

\end{document}

%% file: generated/final_n40_metrics.tex
\newcommand{\FinalPaperCount}{40}
\newcommand{\FinalPosterCount}{120}
\newcommand{\FinalEOneSuccess}{600}
\newcommand{\FinalEOneAttempts}{720}

\newcommand{\FinalEOneRate}{83.3}

\newcommand{\FinalEOneLow}{80.3}
\newcommand{\FinalEOneHigh}{86.2}
\newcommand{\FinalEOneStructural}{89.6}
\newcommand{\FinalEOneScientific}{85.8}
\newcommand{\FinalEOneSpatial}{74.6}
\newcommand{\FinalEOneApproved}{470}
\newcommand{\FinalEOneRevised}{126}
\newcommand{\FinalEOneReplaced}{124}
\newcommand{\FinalETwoDiagnoses}{1109}
\newcommand{\FinalETwoAccepted}{876}
\newcommand{\FinalETwoRejected}{233}
\newcommand{\FinalETwoAssessable}{1100}
\newcommand{\FinalETwoCannotAssess}{9}
\newcommand{\FinalETwoCannotAssessRate}{0.8}
\newcommand{\FinalETwoQuality}{67.2}
\newcommand{\FinalETwoLow}{64.2}
\newcommand{\FinalETwoHigh}{70.2}
\newcommand{\FinalETwoSensitivityQuality}{66.8}
\newcommand{\FinalETwoSensitivityLow}{63.7}
\newcommand{\FinalETwoSensitivityHigh}{69.7}
\newcommand{\FinalEThreeSuccess}{767}
\newcommand{\FinalEThreeAttempts}{876}

\newcommand{\FinalEThreeRate}{87.6}

\newcommand{\FinalEThreeLow}{84.2}
\newcommand{\FinalEThreeHigh}{91.0}
\newcommand{\FinalGeneratedStructuralProposals}{127}
\newcommand{\FinalGeneratedStructuralAccepted}{105}
\newcommand{\FinalExpertStructuralProposals}{2}

%% file: generated/final_completed_primary_metrics.tex
\newcommand{\FinalEFourTargets}{865}
\newcommand{\FinalEFourCannotAssess}{11}
\newcommand{\FinalEFourPosters}{118}
\newcommand{\FinalEFourBefore}{65.0}
\newcommand{\FinalEFourAfter}{87.7}
\newcommand{\FinalEFourUplift}{22.7}
\newcommand{\FinalEFourLow}{17.6}
\newcommand{\FinalEFourHigh}{27.5}
\newcommand{\FinalEFourPositive}{67.1}
\newcommand{\FinalEFourTie}{18.2}
\newcommand{\FinalEFourNegative}{14.8}

%% file: generated/cross_model_metrics.tex
\newcommand{\CrossPapers}{8}
\newcommand{\CrossPosters}{24}
\newcommand{\CrossEOneTasks}{144}
\newcommand{\CrossTerraEOne}{86.8}
\newcommand{\CrossSolEOne}{91.0}
\newcommand{\CrossClaudeEOne}{87.5}
\newcommand{\CrossQwenEOne}{61.8}
\newcommand{\CrossTerraETwo}{72.3}
\newcommand{\CrossSolETwo}{80.2}
\newcommand{\CrossClaudeETwo}{51.5}
\newcommand{\CrossQwenETwo}{59.7}
\newcommand{\CrossTerraEThree}{87.2}
\newcommand{\CrossSolEThree}{91.7}
\newcommand{\CrossClaudeEThree}{73.6}
\newcommand{\CrossQwenEThree}{82.9}
\newcommand{\CrossTerraEFour}{23.3}
\newcommand{\CrossSolEFour}{10.5}
\newcommand{\CrossClaudeEFour}{0.5}
\newcommand{\CrossQwenEFour}{0.9}
\newcommand{\CrossETwoLow}{51.5}
\newcommand{\CrossETwoHigh}{80.2}
\newcommand{\CrossEThreeLow}{73.6}
\newcommand{\CrossEThreeHigh}{91.7}
\newcommand{\CrossEFourLow}{0.5}
\newcommand{\CrossEFourHigh}{23.3}
\newcommand{\CrossClaudeDiagnoses}{368}
\newcommand{\CrossSolDiagnoses}{167}
\newcommand{\CrossClaudeDiagnosesPerPoster}{15.3}
\newcommand{\CrossSolDiagnosesPerPoster}{7.0}
\newcommand{\CrossTargetSpecificity}{3.76}
\newcommand{\CrossActionability}{1.78}
\newcommand{\CrossTargetSpecificityHigh}{92.4}
\newcommand{\CrossActionabilityHigh}{28.2}
\newcommand{\CrossConsensusOneQuality}{28.8}
\newcommand{\CrossConsensusTwoQuality}{38.1}
\newcommand{\CrossConsensusThreeQuality}{61.4}
\newcommand{\CrossConsensusFourQuality}{84.2}
\newcommand{\CrossConsensusOneEThree}{55.1}
\newcommand{\CrossConsensusTwoEThree}{81.0}
\newcommand{\CrossConsensusThreeEThree}{83.8}
\newcommand{\CrossConsensusFourEThree}{87.5}
\newcommand{\CrossConsensusOneEFour}{-7.8}
\newcommand{\CrossConsensusTwoEFour}{-10.0}
\newcommand{\CrossConsensusThreeEFour}{1.2}
\newcommand{\CrossConsensusFourEFour}{18.0}
\newcommand{\CrossPosterGenLow}{21.9}
\newcommand{\CrossPosterGenHigh}{35.0}
\newcommand{\CrossMissingContentUplift}{15.2}
\newcommand{\CrossWhitespaceUplift}{31.2}
\newcommand{\CrossRedundancyUplift}{-21.5}
\newcommand{\CrossCollisionUplift}{-14.8}

%% file: generated/conference_challenge_metrics.tex
\newcommand{\ConferenceEOneSuccess}{96}
\newcommand{\ConferenceEOneAttempts}{240}
\newcommand{\ConferenceEOneRate}{40.0}
\newcommand{\ConferenceETwoDiagnoses}{360}
\newcommand{\ConferenceETwoAssessable}{360}
\newcommand{\ConferenceETwoQuality}{39.8}
\newcommand{\ConferenceETwoLow}{35.1}
\newcommand{\ConferenceETwoHigh}{44.5}
\newcommand{\ConferenceETwoSensitivityQuality}{39.7}
\newcommand{\ConferenceETwoAccepted}{62}
\newcommand{\ConferenceETwoRejected}{298}
\newcommand{\ConferenceETwoAcceptanceRate}{17.2}
\newcommand{\ConferenceEThreeSuccess}{50}
\newcommand{\ConferenceEThreeAttempts}{62}
\newcommand{\ConferenceEThreeRate}{80.6}
\newcommand{\ConferenceEThreeExecutedUnresolved}{11}
\newcommand{\ConferenceEThreeFailed}{1}
\newcommand{\ConferenceEndToEndYield}{13.9}
\newcommand{\ConferenceEFourPairs}{40}
\newcommand{\ConferenceEFourPairsWithTargets}{19}
\newcommand{\ConferenceEFourAssessablePairs}{18}
\newcommand{\ConferenceEFourNoTargetPairs}{21}

\newcommand{\ConferenceEFourTargets}{62}
\newcommand{\ConferenceEFourAssessableTargets}{52}
\newcommand{\ConferenceEFourCannotAssessTargets}{10}
\newcommand{\ConferenceEFourBefore}{82.9}
\newcommand{\ConferenceEFourAfter}{76.7}
\newcommand{\ConferenceEFourUplift}{-6.1}
\newcommand{\ConferenceEFourLow}{-19.3}
\newcommand{\ConferenceEFourHigh}{6.3}
\newcommand{\ConferenceEFourPositive}{19.2}
\newcommand{\ConferenceEFourTie}{61.5}
\newcommand{\ConferenceEFourNegative}{19.2}
\newcommand{\ConferenceStructuredCount}{7}
\newcommand{\ConferenceFreeformCount}{23}
\newcommand{\ConferenceUncertainCount}{10}

\newcommand{\ConferenceFreeformAccepted}{53}
\newcommand{\ConferenceUncertainAccepted}{9}
\newcommand{\ConferenceStructuralDiagnoses}{181}
\newcommand{\ConferenceScientificDiagnoses}{156}
\newcommand{\ConferenceSpatialDiagnoses}{23}

%% file: figures/fig1_system_overview.tex
\begin{figure*}[!t]
\centering
\includegraphics[width=0.90\textwidth]{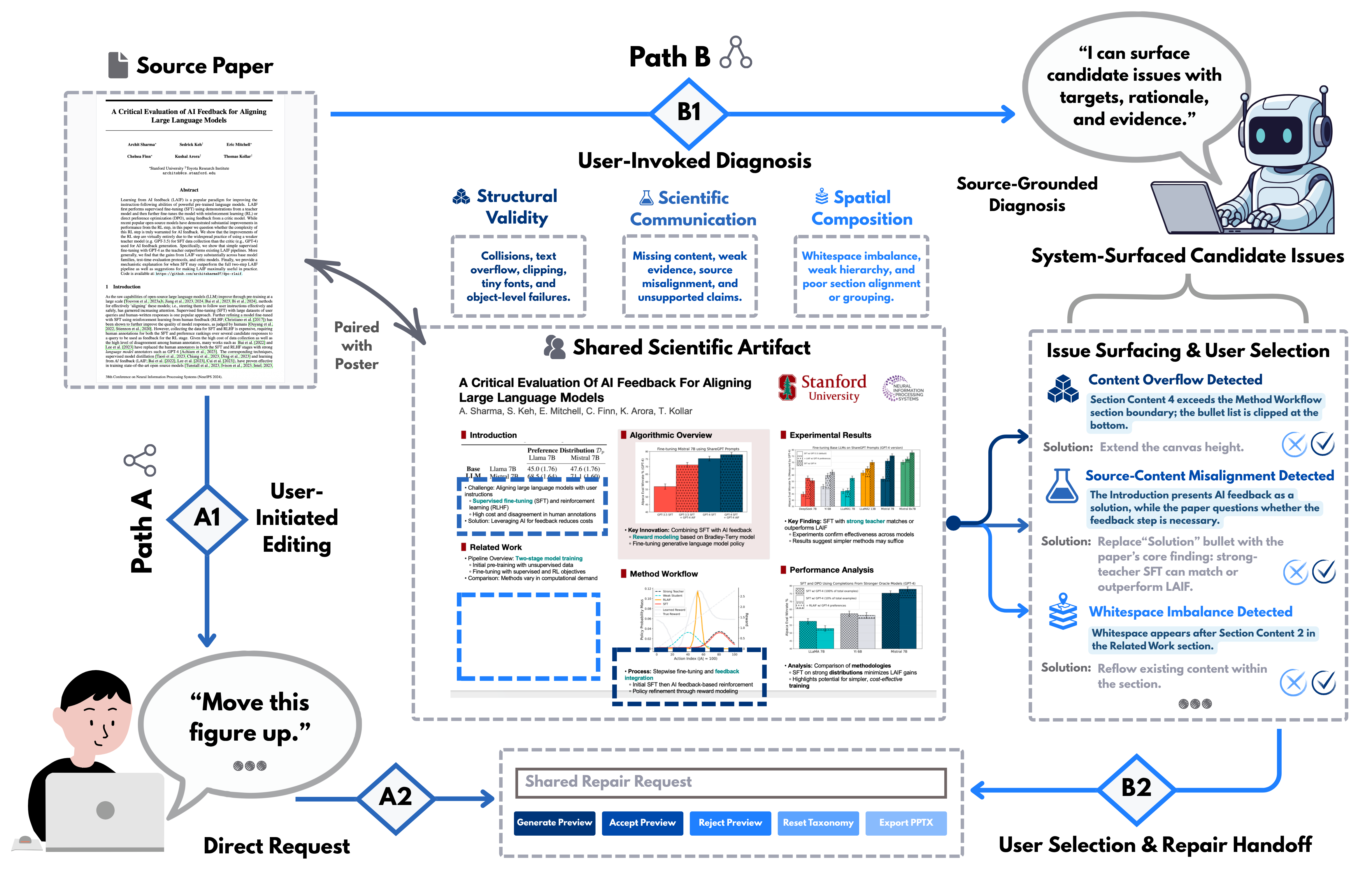}
\caption{\PROS{} separates intent execution from intent formation. Path A begins with an articulated user request. Path B begins before a concrete edit instruction exists: user-invoked diagnosis surfaces source-grounded candidate issues, but only user acceptance turns a candidate into a repair goal. Both paths converge on the same native-object executor, reversible preview, validation, and explicit commitment boundary.}
\Description{A system overview with two entry paths. A source paper is paired with an editable poster. Path A supports a direct user editing request. Path B invokes system-suggested Structural, Scientific, and Spatial diagnoses, surfaces localized candidate issues with rationales and evidence, and lets the user accept or reject each issue. Accepted requests from either path enter the same preview, approval, rejection, and PPTX-export interface.}
\label{fig:overview}
\end{figure*}

%% file: figures/fig2_benchmark.tex
\begin{figure*}[t]
  \centering
  \includegraphics[width=0.90\textwidth]{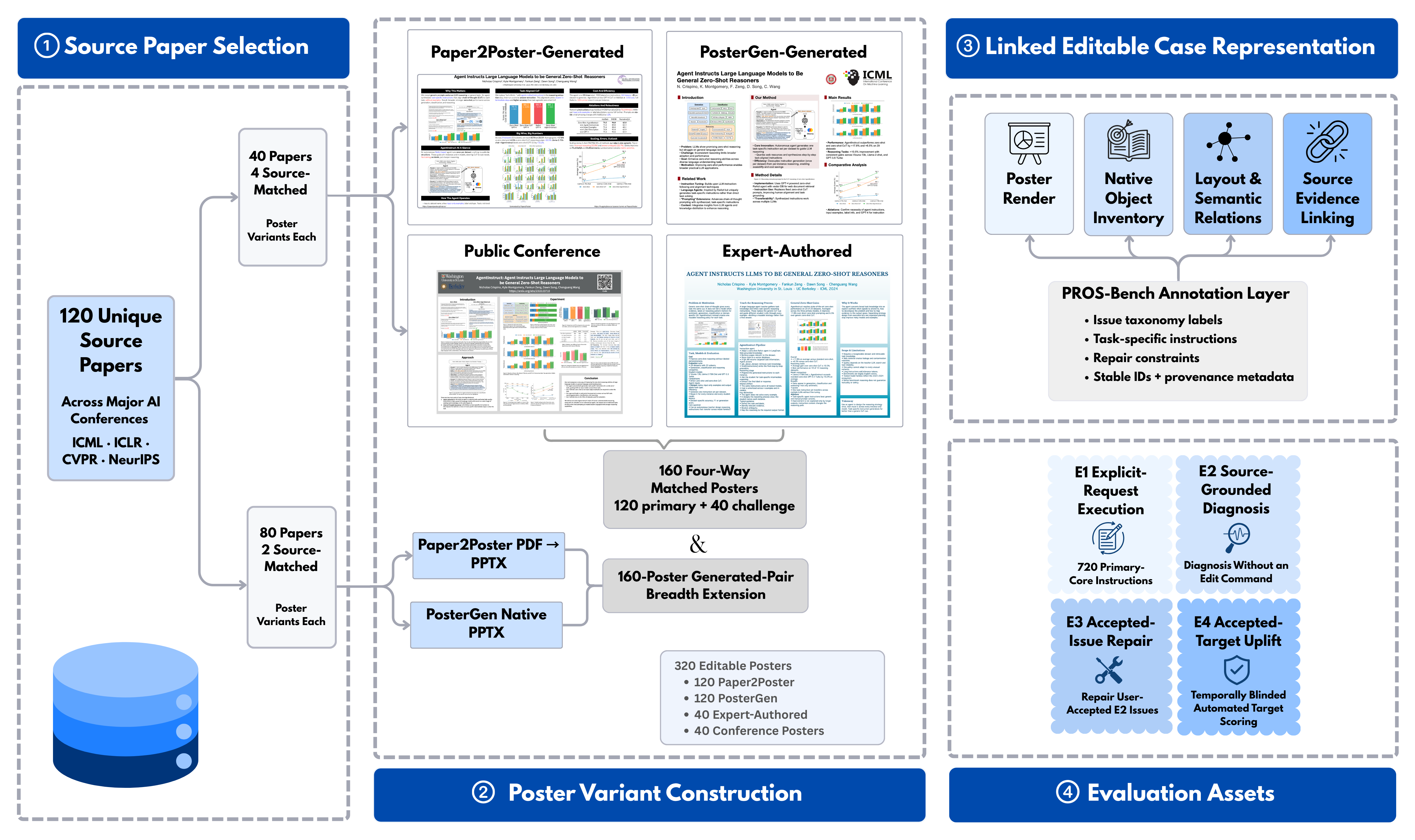}
  \caption{\PROSBench{} links 120 source papers to 320 editable PPTX posters. Forty papers form a 120-poster primary repair core plus a separately reported 40-poster conference representation challenge; 80 additional papers form a 160-poster generated-pair breadth extension. The linked representation preserves renders, native objects, layout and semantic relations, source evidence, stable IDs, provenance, and the E1--E4 evaluation assets.}
  \Description{A four-stage benchmark pipeline. Stage one selects 120 source papers, including 40 four-way paired papers and 80 two-way paired papers. Stage two constructs 320 editable posters spanning Paper2Poster, PosterGen, expert-authored, and original conference variants. Stage three links poster renders, native object inventories, layout and semantic relations, source evidence, and stable provenance metadata. Stage four supplies evaluation assets for reactive editing, proactive diagnosis, grounded repair, and temporally blinded automated accepted-target uplift.}
  \label{fig:benchmark}
\end{figure*}

%% file: figures/fig3_primary_results.tex
\begin{figure*}[t]
\centering
\begin{minipage}[t]{0.38\textwidth}
\centering
\textcolor{resultink}{\sffamily\textbf{(a) Capability-specific outcomes}}\par\vspace{1mm}
\begin{tikzpicture}
\begin{axis}[
  prosplot,
  width=0.93\linewidth,
  height=58mm,
  xmin=0,xmax=100,
  xtick={0,25,50,75,100},
  symbolic y coords={E3 repair,E2 diagnosis,E1 instruction},
  ytick=data,
  xlabel={Capability-specific score (0--100)},
  xmajorgrids,
  enlarge y limits=.32,
  clip=false
]
\draw[resultbluebar,line width=4pt,line cap=round] (axis cs:0,E1 instruction) -- (axis cs:\FinalEOneRate,E1 instruction);
\draw[resultbluebar,line width=4pt,line cap=round] (axis cs:0,E2 diagnosis) -- (axis cs:\FinalETwoQuality,E2 diagnosis);
\draw[resultbluebar,line width=4pt,line cap=round] (axis cs:0,E3 repair) -- (axis cs:\FinalEThreeRate,E3 repair);
\draw[resultblue,line width=3.2pt,line cap=round] (axis cs:\FinalEOneLow,E1 instruction) -- (axis cs:\FinalEOneHigh,E1 instruction);
\draw[resultblue,line width=3.2pt,line cap=round] (axis cs:\FinalETwoLow,E2 diagnosis) -- (axis cs:\FinalETwoHigh,E2 diagnosis);
\draw[resultblue,line width=3.2pt,line cap=round] (axis cs:\FinalEThreeLow,E3 repair) -- (axis cs:\FinalEThreeHigh,E3 repair);
\addplot+[only marks,mark=*,mark size=3.6pt,mark options={draw=resultink,fill=resultblue,line width=.45pt}] coordinates {
  (\FinalEOneRate,E1 instruction)
  (\FinalETwoQuality,E2 diagnosis)
  (\FinalEThreeRate,E3 repair)
};
\node[anchor=west,font=\sffamily\bfseries\scriptsize,text=resultink] at (axis cs:\FinalEOneRate,E1 instruction) {\FinalEOneRate};
\node[anchor=west,font=\sffamily\bfseries\scriptsize,text=resultink] at (axis cs:\FinalETwoQuality,E2 diagnosis) {\FinalETwoQuality};
\node[anchor=west,font=\sffamily\bfseries\scriptsize,text=resultink] at (axis cs:\FinalEThreeRate,E3 repair) {\FinalEThreeRate};
\end{axis}
\end{tikzpicture}
\end{minipage}\hfill
\begin{minipage}[t]{0.58\textwidth}
\centering
\textcolor{resultink}{\sffamily\textbf{(b) Paper-level accepted-target uplift}}\par\vspace{1mm}
\pgfplotstableread{generated/e4_paper_level_pairs.dat}\eFourPaperPairs
\begin{tikzpicture}
\begin{axis}[
  prosplot,
  width=0.86\linewidth,
  height=58mm,
  xmin=-20,xmax=55,
  xtick={-20,0,20,40},
  ymin=-0.48,ymax=0.48,
  ytick=\empty,
  xlabel={Paper-level target-condition uplift (percentage points)},
  xmajorgrids,
  clip=false
]
\draw[resultcoral,dashed,line width=.7pt] (axis cs:0,-.43) -- (axis cs:0,.43);
\addplot[only marks,mark=*,mark size=1.4pt,opacity=1,mark options={draw=resultblue,fill=resultblue,line width=.22pt}]
  table[x expr=\thisrow{vlm_after}-\thisrow{vlm_before},y expr={0.055*(mod(\coordindex,9)-4)}] {generated/e4_paper_level_pairs.dat};
\draw[resultbluebar,line width=3pt,line cap=round] (axis cs:\FinalEFourLow,0) -- (axis cs:\FinalEFourHigh,0);
\addplot[only marks,mark=diamond*,mark size=3.6pt,mark options={draw=resultink,fill=resultblue,line width=.5pt}] coordinates {(\FinalEFourUplift,0)};
\node[anchor=south west,font=\sffamily\bfseries\scriptsize,text=resultblue,fill=white,inner sep=1pt] at (axis cs:\FinalEFourUplift,.29) {mean +\FinalEFourUplift{} pp};
\node[anchor=north east,font=\sffamily\scriptsize,text=resultink!72] at (axis cs:53,-.28) {40 source papers};
\end{axis}
\end{tikzpicture}
\end{minipage}

\vspace{2mm}
\begin{minipage}{0.96\textwidth}
\centering
\textcolor{resultink}{\sffamily\textbf{(c) Accepted-target outcome distribution}}\par\vspace{0.5mm}
\begin{tikzpicture}
\begin{axis}[
  prosplot,
  width=0.80\linewidth,
  height=32mm,
  xmin=0,xmax=75,
  xtick={0,20,40,60},
  symbolic y coords={Declined,Tied,Improved},
  ytick={Declined,Tied,Improved},
  yticklabels={Declined,Tied,Improved},
  xlabel={Share of accepted repair targets (\%)},
  xmajorgrids,
  enlarge y limits=.28,
  clip=false
]
\draw[resultbluebar,line width=3pt,line cap=round] (axis cs:0,Improved) -- (axis cs:\FinalEFourPositive,Improved);
\draw[resultnavy,opacity=.55,line width=3pt,line cap=round] (axis cs:0,Tied) -- (axis cs:\FinalEFourTie,Tied);
\draw[resultredbar,line width=3pt,line cap=round] (axis cs:0,Declined) -- (axis cs:\FinalEFourNegative,Declined);
\addplot[only marks,mark=*,mark size=3.5pt,mark options={draw=resultink,fill=resultblue,line width=.45pt}] coordinates {(\FinalEFourPositive,Improved)};
\addplot[only marks,mark=*,mark size=3.5pt,mark options={draw=resultink,fill=resultnavy,line width=.45pt}] coordinates {(\FinalEFourTie,Tied)};
\addplot[only marks,mark=*,mark size=3.5pt,mark options={draw=resultink,fill=resultcoral,line width=.45pt}] coordinates {(\FinalEFourNegative,Declined)};
\node[anchor=west,font=\sffamily\bfseries\scriptsize,text=resultblue] at (axis cs:\FinalEFourPositive,Improved) {\FinalEFourPositive\%};
\node[anchor=west,font=\sffamily\bfseries\scriptsize,text=resultnavy] at (axis cs:\FinalEFourTie,Tied) {\FinalEFourTie\%};
\node[anchor=west,font=\sffamily\bfseries\scriptsize,text=resultcoral!80!black] at (axis cs:\FinalEFourNegative,Declined) {\FinalEFourNegative\%};
\end{axis}
\end{tikzpicture}
\end{minipage}
\caption{Primary-core outcome profile over 40 papers and 120 posters. (a) E1 verified instruction success, E2 stage-balanced diagnosis quality, and E3 verified target-resolution success remain separate capability-specific measures; intervals are paper-clustered 95\% confidence intervals. (b) Each point is one paper's stage- and poster-balanced uplift under the temporally blinded automated VLM evaluation. The diamond is the paper-macro mean and the thick segment is its 95\% paper-clustered confidence interval. (c) Target-level outcomes retain improved, tied, and declined cases rather than implying a uniform shift. Together, the panels show why verified local resolution and realized final-state benefit should not be treated as equivalent outcomes; no panel is a composite system score.}
\Description{A three-panel results figure. Panel A shows E1 instruction success at 83.3, E2 diagnosis quality at 67.2, and E3 repair success at 87.6, with confidence intervals. Panel B is a strip plot of automated VLM paper-level uplift for 40 source papers, spanning declines through gains of more than 40 points, with a mean diamond at 22.7 points and a 95 percent confidence interval. Panel C uses three horizontal lollipops to show 67.1 percent of accepted targets improving, 18.2 percent tying, and 14.8 percent declining.}
\label{fig:primary-results}
\end{figure*}
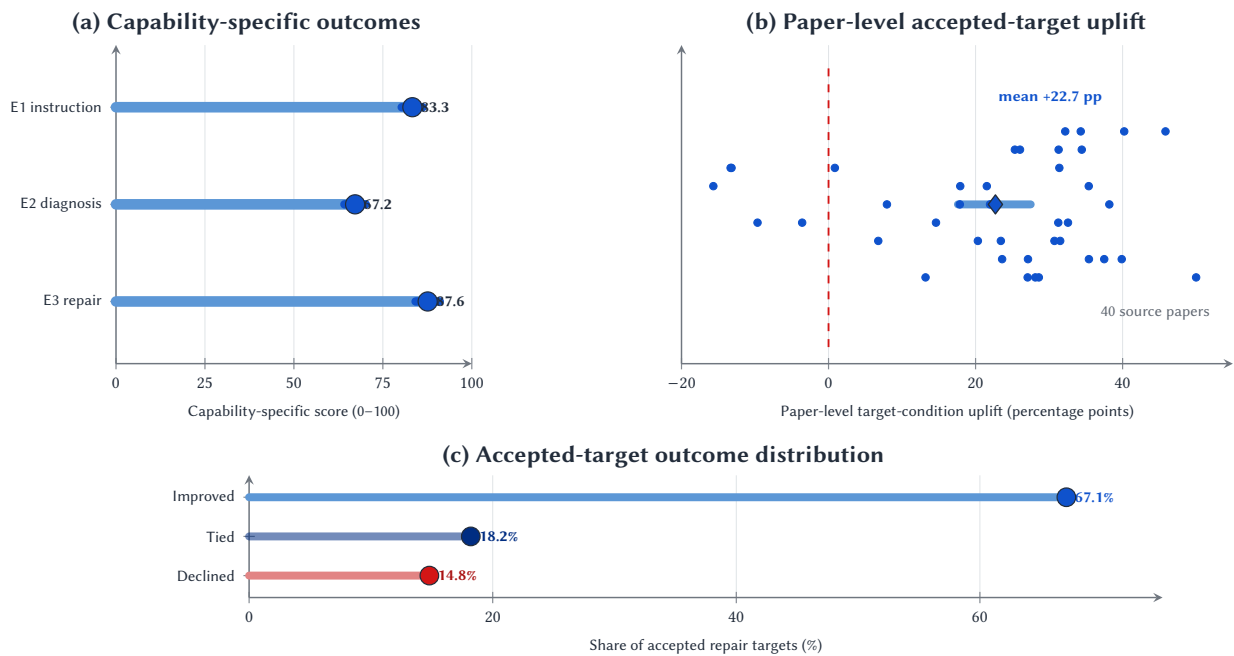

%% file: figures/fig5_qualitative_upshift.tex
\begin{figure*}[t]
  \centering
  \includegraphics[width=0.90\textwidth]{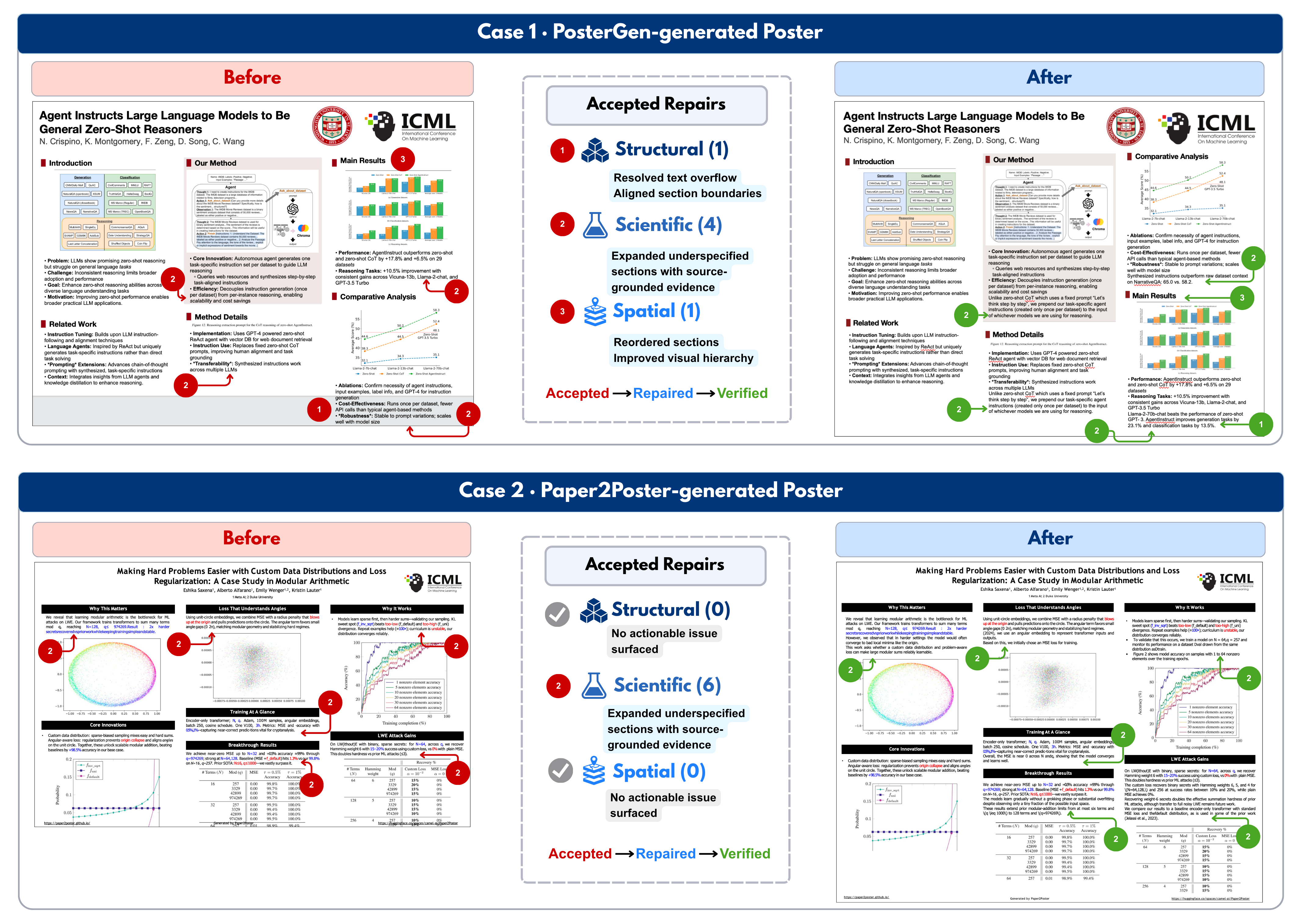}
  \caption{Two qualitative before/after repair traces. In the PosterGen case, accepted Structural, Scientific, and Spatial repairs resolve overflow, expand underspecified sections with source-grounded evidence, reorder sections, and strengthen visual hierarchy. In the Paper2Poster case, accepted Scientific repairs expand underspecified sections with source-grounded evidence; no actionable Structural or Spatial issue is surfaced. Red callouts mark accepted targets before repair, green callouts identify the corresponding regions after repair, and numerals denote the diagnosis stage rather than the number of repairs. The after-state is produced only after user acceptance, repair, and verification.}
  \Description{Two qualitative cases. Each case shows a before poster, a center panel summarizing repairs by stage, and an after poster. Red numbered callouts indicate accepted target regions in the before posters, and green numbered callouts indicate corresponding regions in the after posters. In the second case, neutral gray checks mark stages in which no actionable issue was surfaced.}
  \label{fig:qualitative}
\end{figure*}

%% file: figures/fig5_calibration_findings.tex
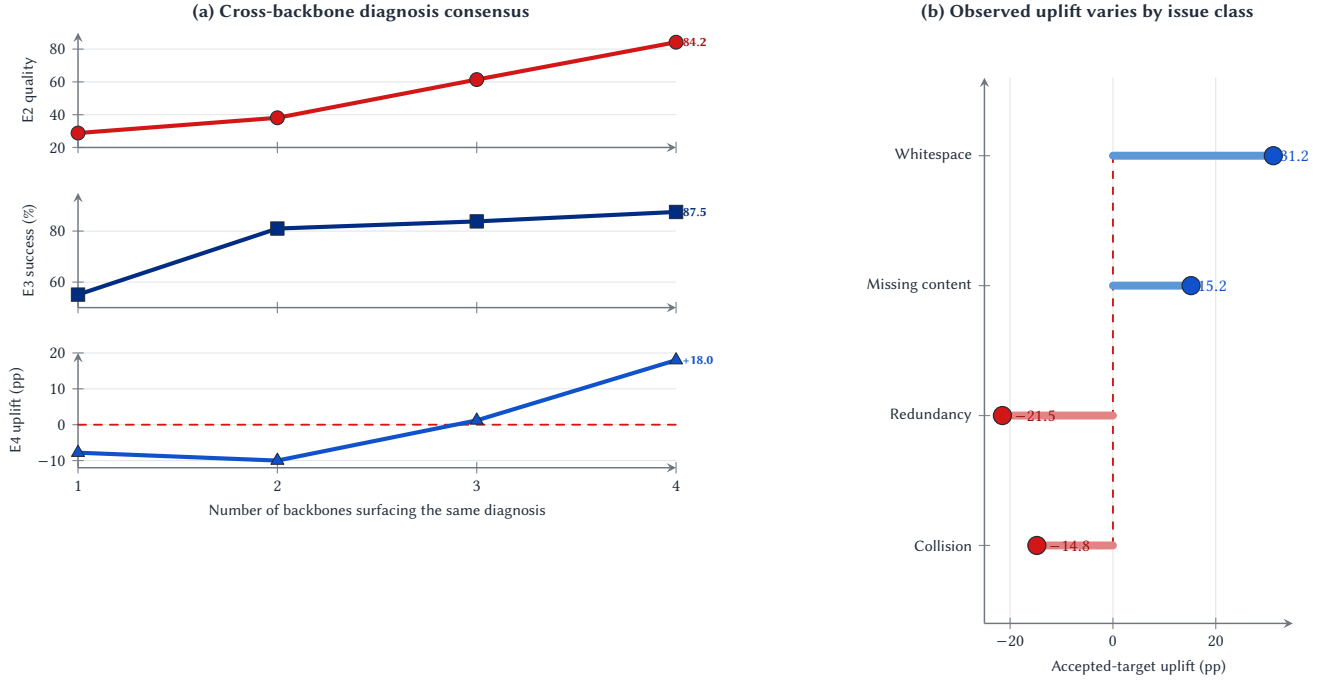
\begin{figure*}[t]
\centering
\begin{minipage}[t]{0.55\textwidth}
\centering
\textcolor{resultink}{\sffamily\textbf{(a) Cross-backbone diagnosis consensus}}\par\vspace{1mm}
\begin{tikzpicture}
\begin{groupplot}[
  prosplot,
  group style={group size=1 by 3,vertical sep=6mm},
  width=0.97\linewidth,
  height=31mm,
  xmin=1,xmax=4,
  xtick={1,2,3,4},
  ymajorgrids,
  clip=false
]
\nextgroupplot[ymin=20,ymax=90,ylabel={E2 quality},xticklabels={}]
\addplot+[line width=1.55pt,resultcoral,mark=*,mark size=2.6pt,mark options={draw=resultink,fill=resultcoral,line width=.4pt}] coordinates {(1,\CrossConsensusOneQuality) (2,\CrossConsensusTwoQuality) (3,\CrossConsensusThreeQuality) (4,\CrossConsensusFourQuality)};
\node[anchor=west,font=\sffamily\bfseries\tiny,text=resultcoral!82!black] at (axis cs:4,\CrossConsensusFourQuality) {\CrossConsensusFourQuality};
\nextgroupplot[ymin=50,ymax=95,ylabel={E3 success (\%)},xticklabels={}]
\addplot+[line width=1.55pt,resultnavy,mark=square*,mark size=2.5pt,mark options={draw=resultink,fill=resultnavy,line width=.4pt}] coordinates {(1,\CrossConsensusOneEThree) (2,\CrossConsensusTwoEThree) (3,\CrossConsensusThreeEThree) (4,\CrossConsensusFourEThree)};
\node[anchor=west,font=\sffamily\bfseries\tiny,text=resultnavy] at (axis cs:4,\CrossConsensusFourEThree) {\CrossConsensusFourEThree};
\nextgroupplot[ymin=-12,ymax=20,ylabel={E4 uplift (pp)},xlabel={Number of backbones surfacing the same diagnosis}]
\addplot[resultcoral,dashed,line width=.7pt] coordinates {(1,0) (4,0)};
\addplot+[line width=1.55pt,resultblue,mark=triangle*,mark size=2.8pt,mark options={draw=resultink,fill=resultblue,line width=.4pt}] coordinates {(1,\CrossConsensusOneEFour) (2,\CrossConsensusTwoEFour) (3,\CrossConsensusThreeEFour) (4,\CrossConsensusFourEFour)};
\node[anchor=west,font=\sffamily\bfseries\tiny,text=resultblue] at (axis cs:4,\CrossConsensusFourEFour) {+\CrossConsensusFourEFour};
\end{groupplot}
\end{tikzpicture}
\end{minipage}\hfill
\begin{minipage}[t]{0.37\textwidth}
\centering
\textcolor{resultink}{\sffamily\textbf{(b) Observed uplift varies by issue class}}\par\vspace{7mm}
\begin{tikzpicture}
\begin{axis}[
  prosplot,
  width=0.86\linewidth,
  height=88mm,
  xmin=-25,xmax=35,
  symbolic y coords={Collision,Redundancy,Missing content,Whitespace},
  ytick={Collision,Redundancy,Missing content,Whitespace},
  xlabel={Accepted-target uplift (pp)},
  xmajorgrids,
  enlarge y limits=.20,
  clip=false
]
\draw[resultcoral,dashed,line width=.7pt] (axis cs:0,Collision) -- (axis cs:0,Whitespace);
\draw[resultbluebar,line width=3pt,line cap=round] (axis cs:0,Missing content) -- (axis cs:\CrossMissingContentUplift,Missing content);
\draw[resultbluebar,line width=3pt,line cap=round] (axis cs:0,Whitespace) -- (axis cs:\CrossWhitespaceUplift,Whitespace);
\draw[resultredbar,line width=3pt,line cap=round] (axis cs:0,Redundancy) -- (axis cs:\CrossRedundancyUplift,Redundancy);
\draw[resultredbar,line width=3pt,line cap=round] (axis cs:0,Collision) -- (axis cs:\CrossCollisionUplift,Collision);
\addplot+[only marks,mark=*,mark size=3.4pt,mark options={draw=resultink,fill=resultblue,line width=.45pt},nodes near coords,nodes near coords style={font=\sffamily\bfseries\scriptsize,text=resultblue,anchor=west},point meta=x] coordinates {(\CrossMissingContentUplift,Missing content) (\CrossWhitespaceUplift,Whitespace)};
\addplot+[only marks,mark=*,mark size=3.4pt,mark options={draw=resultink,fill=resultcoral,line width=.45pt},nodes near coords,nodes near coords style={font=\sffamily\bfseries\scriptsize,text=resultcoral!75!black,anchor=west,xshift=2pt},point meta=x] coordinates {(\CrossRedundancyUplift,Redundancy) (\CrossCollisionUplift,Collision)};
\end{axis}
\end{tikzpicture}
\end{minipage}
\caption{Descriptive post-hoc patterns on the fixed eight-paper cross-model subset. (a) Mean E2 quality and E3 success increase across agreement bins, whereas E4 is negative at one and two backbones, near zero at three, and positive at four. Shared prompts, detectors, artifact state, and judge mean that agreement is neither statistically independent nor an evaluated ensemble. (b) Observed accepted-target uplift differs across issue classes, but these means also combine model, artifact, target-selection, baseline-headroom, and trajectory differences. Both panels are hypothesis-generating and do not report inferential correlations or causal effects.}
\Description{Two panels summarize exploratory post-hoc patterns. The left panel contains three aligned line charts: red for diagnosis quality, dark blue for repair success, and blue for targeted uplift as the number of backbones in an agreement bin rises from one to four. Diagnosis quality and repair success rise across the bins, while targeted uplift is negative for one and two backbones, near zero for three, and positive for four. The right panel is a diverging lollipop chart: blue missing-content and whitespace targets have positive observed mean uplift, while red redundancy and collision targets have negative observed mean uplift.}
\label{fig:calibration-findings}
\end{figure*}